\documentclass[sn-mathphys-ay]{sn-jnl}

\usepackage{graphicx}%
\usepackage{pdflscape}%
\usepackage{array}%
\usepackage{multirow}%
\usepackage{amsmath,amssymb,amsfonts}%
\usepackage{amsthm}%
\usepackage{mathrsfs}%
\usepackage[title]{appendix}%
\usepackage{xcolor}%
\usepackage{textcomp}%
\usepackage{manyfoot}%
\usepackage{booktabs}%
\usepackage{algorithm}%
\usepackage{algorithmicx}%
\usepackage{algpseudocode}%
\usepackage{listings}%

\newcommand\degr{\ensuremath{^\circ}}

\begin{document}

\title[Accreting Neutron Stars in Low-Mass X-ray Binaries]{Accreting Neutron Stars in Low-Mass X-ray Binaries}


\author*[1]{\fnm{M. Cristina} \sur{Baglio}}\email{maria.baglio@inaf.it}

\author[1]{\fnm{Sara E.} \sur{Motta}}\email{sara.motta@inaf.it}

\author[2]{\fnm{Anagha P.} \sur{Nitindala}}\email{apniti@utu.fi}

\author[2]{\fnm{Juri} \sur{Poutanen}}\email{juri.poutanen@utu.fi}

\affil*[1]{\orgdiv{INAF--Osservatorio Astronomico di Brera}, \orgname{INAF}, \orgaddress{\street{Via Bianchi 46}, \city{Merate}, \postcode{I-23807}, \country{Italy}}}

\affil[2]{\orgdiv{Department of Physics and Astronomy}, \orgname{University of Turku}, \orgaddress{\postcode{FI-20014}, \country{Finland}}}


\abstract{Neutron star low-mass X-ray binaries (NS LMXBs) are systems in which a neutron star accretes matter from a Roche-lobe-filling companion via an accretion disc, releasing gravitational energy with an efficiency of order 10--20\% of the rest-mass energy. Unlike black holes, neutron stars possess a solid surface and a magnetic field, making NS LMXBs uniquely rich laboratories for high-energy astrophysics and dense-matter physics. In this article we review the current understanding of accretion onto neutron stars in LMXBs, tracing matter from the outer disc radii to the stellar surface across five decades of observational discovery. We survey the principal classes of accreting neutron star systems --- canonical atoll and Z-sources, ultracompact X-ray binaries, accreting millisecond pulsars, and transitional millisecond pulsars --- and discuss how their spectral states, timing properties, and multiwavelength behaviour encode the physics of the inner accretion flow. We examine thermonuclear Type-I X-ray bursts as probes of nuclear burning and neutron star structure, and review the boundary and spreading layers where accreted matter decelerates onto the stellar surface. We discuss relativistic jets and ionised disc winds as regulators of the accretion--ejection cycle. Finally, we present X-ray polarimetric results from the \textit{Imaging X-ray Polarimetry Explorer} (\textit{IXPE}), which for the first time provide new, though model-dependent, constraints on the geometry of the emission regions --- disc, corona, spreading layer, and jets --- that spectroscopy alone cannot resolve, alongside complementary optical and infrared polarimetric observations.}

\keywords{accretion discs, neutron star, X-ray binary, X-ray bursts, millisecond pulsars}



\maketitle


\section{Introduction}\label{sec1}

Neutron star (NS) low-mass X-ray binaries (LMXBs) are systems in which a NS accretes matter from a low-mass companion star via Roche lobe overflow. 
The infalling material forms an accretion disc and ultimately reaches the stellar surface, releasing gravitational energy with an efficiency of order $\sim$10--20\% of the rest-mass energy, making NS~LMXBs among the brightest persistent and transient sources in the X-ray sky. 
These systems serve as astrophysical laboratories of unique value: the accretion disc, the boundary and spreading layers where Keplerian flow decelerates onto the stellar surface, the thermonuclear bursts ignited in the accreted envelope, and the jets and winds driven off the disc are all diagnostics that encode, across a broad range of wavelengths, information about the mass, radius, spin, and magnetic field of the underlying NS. 
Unlike black holes (BHs), NSs possess a solid surface and a magnetic field --- and it is precisely these features that make the phenomenology of NS~LMXBs qualitatively richer and in many respects more constraining than that of their BH counterparts.

The study of NS~LMXBs has been shaped by successive observational advances over the past fifty years. 
The recycling scenario --- in which old, weakly magnetised NSs are spun up to millisecond periods by the transfer of angular momentum during accretion --- has its roots in the 1970s \citep{BisnovatyiKogan1974,SmarrBlandford1976}, the latter based on the anomalously weak field and fast spin of the binary radio pulsar PSR~B1913+16, and was developed into a formation channel for millisecond pulsars following the discovery of the first MSP, PSR~B1937+21 \citep{Backer1982,Alpar82,Radhakrishnan1982}.

The mapping of spectral states and colour-colour diagrams with \textit{EXOSAT} and \textit{Ginga} established the canonical atoll and Z-source classification and linked it to changes in the accretion rate and inner-flow geometry \citep{Hasinger1989}.
The subsequent discovery with \textit{RXTE} of coherent X-ray pulsations with millisecond periods in persistent emission of transiently accreting NSs \citep{Wijnands98} and during thermonuclear bursts \citep{Strohmayer1996}, together with the later identification of transitional millisecond pulsars (Sect.~\ref{tmsp_sec}), did not establish the recycling scenario but rather confirmed it observationally, filling in the missing evolutionary links between accreting and rotation-powered millisecond pulsars.
Broadband spectroscopy with \textit{XMM-Newton}, \textit{Chandra}, and \textit{NuSTAR} revealed relativistic reflection features \citep{Cackett2010} and disc winds \citep{DiazTrigoBoirin2016} as ubiquitous components of the accretion flow, while coordinated radio, optical, and infrared campaigns established the connection between accretion state and jet activity \citep{Migliari2006}, and placed NS~LMXBs firmly in the broader context of the disc--jet coupling observed across the black-hole binary population \citep{Gallo2003}.

Most recently, the \textit{Imaging X-ray Polarimetry Explorer} (\textit{IXPE}; \citealt{Weisskopf2022}) has opened a new observational window, providing the first sensitive measurements of X-ray polarisation in these systems and offering new, model-dependent constraints on the geometry of the emission regions that spectroscopy alone cannot supply.

In this article, we review the current understanding of accretion onto NS in LMXBs, tracing the physical processes from the outer disc to the stellar surface and connecting them to the diverse observational phenomenology that has accumulated over five decades. 
We begin by surveying in Section~\ref{sec:flavours} the principal classes of NS~LMXBs --- canonical atoll and Z-sources, ultracompact X-ray binaries, accreting millisecond pulsars, and transitional millisecond pulsars --- highlighting the physical parameters that govern their behaviour. 
We then examine the physics of accretion itself: the disc structure and spectral states (Section~\ref{sec:variab}), 
the thermonuclear Type-I X-ray bursts that probe nuclear burning under extreme conditions and constrain the NS equation of state (Section~\ref{sec:bursts}), 
and the boundary and spreading layers where accreted matter settles onto the stellar surface (Section~\ref{sec:bl}). 
We follow matter that does not reach the surface, reviewing the relativistic jets (Section~\ref{sec:jets}) and ionised disc winds (Section~\ref{sec:winds}) that regulate the accretion--ejection cycle. 
Finally, in Section~\ref{sec:polarisation} we discuss the polarimetric properties of NS~LMXBs: we start by presenting in Section~\ref{sec:ixpe} new results from \textit{IXPE}, which have begun to resolve long-standing degeneracies in spectral modelling and to constrain the geometry of the accretion flow in ways that were not previously accessible, and then in Section~\ref{sec:optir_pol} we discuss polarimetric observations at optical and infrared wavelengths. 

Throughout, the NS surface plays a central role --- it is the feature that most fundamentally distinguishes NS~LMXBs from black-hole X-ray binaries, and it is the reason why these systems remain, more than sixty years after their discovery, one of the richest laboratories for high-energy astrophysics and dense-matter physics.

\section{The different flavours of accreting NSs}\label{sec:flavours}

NS LMXBs display a wide phenomenological diversity, reflecting the variety of physical conditions under which accretion onto a NS can occur. While the canonical atoll and Z sources primarily trace different accretion regimes at varying mass accretion rates (Sect. \ref{Canonical_NS}), additional classes of systems highlight the role of other key parameters, such as the nature of the donor star, the strength of the NS magnetic field, and the geometry of the inner accretion flow. These include ultra-compact X-ray binaries (UCXBs; Sect. \ref{UCXBs}), in which hydrogen-deficient donors and short orbital periods lead to distinct accretion environments, and accreting millisecond X-ray pulsars (AMXPs; Sect. \ref{sec:amxp}), where the magnetic field is sufficiently strong to channel the accretion flow onto the magnetic poles, producing coherent pulsations. 
Together, these systems illustrate the many ``flavours'' of accretion onto NSs, and provide complementary insights into how accretion proceeds under different physical regimes.

\subsection{Canonical NS LMXBs}\label{Canonical_NS}
LMXBs are traditionally divided into two main phenomenological classes, the so-called
\textit{atoll} and \textit{Z} sources, according to the tracks they describe in X-ray
colour--colour diagrams and their associated timing behaviour \citep{Hasinger1989}.
Although this classification was originally empirical, it is now understood to reflect
different accretion regimes and, more generally, different configurations of the inner
accretion flow.

\begin{figure}
    \centering
    \includegraphics[width=0.7\linewidth, angle=-90]{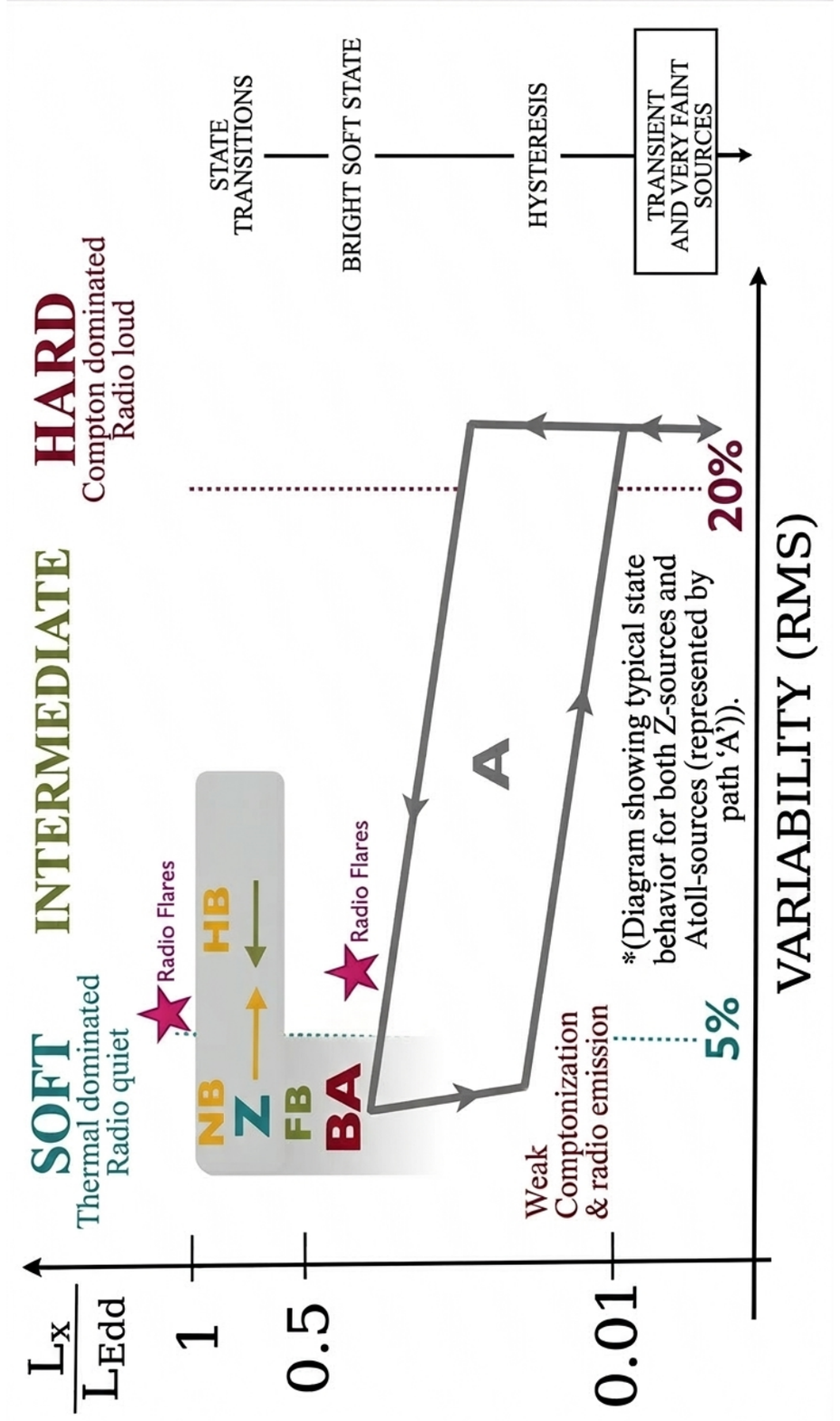}
    \caption{Qualitative behaviour of atoll (A), bright atoll (BA) and Z-sources in
    the RMS-Intensity Diagram (RID); adapted from \cite{MunozDarias2014}.}
    \label{fig:RID}
\end{figure}

Atoll sources are typically observed over a broad range of luminosities, from relatively faint ($L_{\rm X}\sim10^{36}$\,erg\,s$^{-1}$) hard states to bright ($L_{\rm X}\gtrsim10^{37}$\,erg\,s$^{-1}$) soft states. In colour--colour and hardness--intensity diagrams they trace the well-known island and banana branches (see Fig.~\ref{fig:RID}). The island state corresponds to low accretion rates and is characterized by hard spectra (more flux $>10$ keV than below) dominated by Comptonized emission from a hot inner flow, often accompanied by strong band-limited noise and high fractional variability. As the accretion rate increases, sources move into the banana branch, where the spectrum progressively softens (more flux $<10$ keV than above) and becomes dominated by emission from the accretion disc and the boundary/spreading layer. Along the banana branch, the fractional variability decreases and the power spectrum becomes less dominated by a broad-band noise, reflecting a more stable, optically thick inner accretion flow. During outburst
transitions between the hard island state and the soft banana branch, atoll sources
pass through an intermediate state (IS), characterised by intermediate spectral
hardness and variability properties, analogous to the intermediate states observed
in black-hole X-ray binaries \citep[see][in this collection for a review]{Marcel2026}.

Their behaviour is now understood within the same accretion-state framework used
for black-hole X-ray binaries: at low accretion rate the inner disc is likely truncated
and replaced by a geometrically thick, optically thin flow, whereas at higher accretion
rates the disc extends closer to the NS surface and the boundary/spreading layer
becomes increasingly prominent \citep{Done2007,Lin2007}. During outbursts, several
atoll sources also show clear hysteresis patterns analogous to those observed in
black-hole systems \citep{MaccaroneCoppi2003}.

Z-sources, on the other hand, are persistently bright systems accreting close to or above the Eddington limit. In colour--colour diagrams they trace the characteristic horizontal (HB), normal (NB), and flaring (FB) branches, connected by the hard apex (HA) between the HB and NB, and the soft apex (SA) between the NB and FB \citep{Hasinger1989,vanderKlis2006}. In contrast to atolls, they generally remain in soft, optically thick accretion regimes, with comparatively low fractional rms variability \citep{MunozDarias2014}. Their spectra are expected to be dominated by the accretion disc and by a luminous boundary/spreading layer, which at these high accretion rates may cover a substantial fraction of the NS surface \citep{Inogamov1999,Lin2009}.

A more unified picture has emerged in recent years from the combined use of colour,
hardness, luminosity, and variability diagnostics \citep{Belloni2010}. In particular, the rms--intensity
diagram shows that NS LMXBs populate a continuous accretion-state space, with
atolls evolving from hard/high-rms to soft/low-rms states, and Z-sources occupying
the luminous, low-rms extreme of this distribution \citep[Fig.~\ref{fig:RID};][]{MunozDarias2014}.
In this framework, the main driver is the mass accretion rate and its effect on the
geometry of the inner flow, rather than a strict dichotomy between source classes.
A special subset of high-inclination systems, known as accretion-disc corona
(ADC) sources, are viewed at angles where the inner accretion flow is obscured by
the outer disc, so that the observed X-ray emission is dominated by radiation
scattered or reprocessed in an extended corona above the disc plane \citep{WhiteHolt1982}; their
polarimetric properties are discussed in Section~\ref{sec:polarisation}.

This interpretation is strongly supported by sources that bridge the two
phenomenologies. The clearest example is XTE~J1701$-$462, which during its
2006--2007 outburst evolved from Cyg-like Z behaviour to Sco-like Z behaviour and
eventually into an atoll source as the luminosity decreased, demonstrating that the
atoll and Z classes can arise within the same system under changing accretion
conditions \citep{Homan2007,Lin2009}. Canonical atoll and Z sources should
therefore be regarded as different manifestations of accretion onto weakly magnetized
NSs across different luminosity and inner-flow regimes, rather than as fundamentally
distinct source populations.

\subsection{Ultracompact X-Ray binaries}\label{UCXBs}


Ultra-compact X-ray binaries (UCXBs) are a subclass of LMXBs characterized by orbital periods shorter than $\sim 60$--$80$ minutes and hydrogen-deficient donor stars. In these systems, the donor is typically an electron-degenerate or semi-degenerate object, and the material transferred onto the NS is therefore hydrogen-poor, often helium- or carbon/oxygen-rich \citep{Paczynski1981,Rappaport1982}. The chemical composition of the accreted material fundamentally alters the physics of accretion compared to hydrogen-rich LMXBs, affecting the structure and stability of the accretion disc, the thermonuclear burning regime on the NS surface, and the long-term mass-transfer evolution.

In UCXBs, mass transfer is primarily driven by angular momentum losses via gravitational-wave radiation. As a result, the secular mass-transfer rate is tightly coupled to the orbital evolution, providing a relatively clean framework to study accretion under well-defined angular-momentum-loss prescriptions. The small orbital separation implies a correspondingly small accretion disc, which strongly influences the critical accretion rate required for thermal-viscous stability \citep{vanParadijs1994,Lasota2001}. In particular, the threshold for maintaining the disc in a fully ionized state depends sensitively on both disc size and chemical composition, leading to systematic differences in persistent versus transient behaviour compared to wider, hydrogen-rich LMXBs.

\begin{figure} \centering \includegraphics[width=1\linewidth]{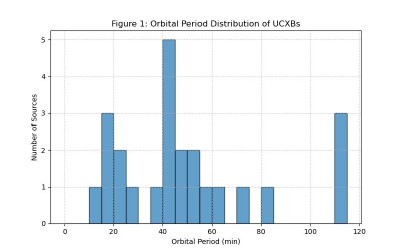} \caption{Histogram showing the distribution of confirmed and candidate UCXBs as a function of their orbital period (built using the data from UltraCompCAT). The two populations of transient ($P_{\rm orb} < 30 \rm \, min$) and persistent ($40 \,\mbox{min} < P_{\rm orb} < 60\,\mbox{min}$) UCXBs can be identified.} 
\label{Fig:ucxb} 
\end{figure}

A comprehensive census of confirmed and candidate UCXBs is provided by UltraCompCAT\footnote{UltraCompCAT is accessible at: \url{https://research.iac.es/proyecto/compactos/UltraCompCAT/}} \citep{ArmasPadilla2023}. The catalogue currently includes 20 confirmed UCXBs with measured orbital periods shorter than 80 minutes, along with additional candidate systems identified through their multi-wavelength accretion signatures. The orbital-period distribution (Fig.~\ref{Fig:ucxb}) reveals two main groups: short-period systems ($P_{\rm orb} \lesssim 30$~min), predominantly persistent, and systems with $40 \, \mbox{min} \lesssim P_{\rm orb} \lesssim 60 \, \mbox{min}$, which are largely transient. This dichotomy reflects differences in secular mass-transfer rates and disc stability thresholds, rather than solely distinct evolutionary origins.

\subsubsection{Evolutionary Pathways}

Several formation channels can lead to the development of UCXBs \citep{Nelemans2010,ArmasPadilla2023}. Although they originate from different progenitor configurations, from an accretion perspective they all converge to Roche-lobe overflow from a hydrogen-deficient donor onto a NS:

\begin{itemize}

\item White Dwarf (WD) Channel: A NS paired with a white dwarf undergoes orbital decay driven by gravitational-wave emission. When the WD fills its Roche lobe, mass transfer begins at relatively high rates in a compact configuration, producing a persistently accreting system. As the orbit subsequently expands and the donor mass decreases, the mass-transfer rate declines, often driving a transition to transient behaviour \citep{Yungelson2002,Sengar2017,vanHaaften2012b}.

\item Helium Star Channel: In this scenario, the donor is a low-mass helium-burning star that initiates mass transfer before becoming fully degenerate. The evolving mass--radius relation of the donor regulates the mass-transfer rate and its long-term evolution, thereby influencing disc stability and the accretion luminosity \citep{Savonije1986,Iben1987,Yungelson2008,Nelemans2010}.

\item Evolved Main-Sequence Channel: Here, mass transfer begins after central hydrogen depletion in an evolved main-sequence star. Angular-momentum losses due to gravitational radiation and magnetic braking drive the orbit to ultra-short periods ($\sim 10$ minutes). The system then follows an evolutionary path similar to the helium star channel, ultimately producing a helium-rich donor \citep{Tutukov1985,Nelson1986,Podsiadlowski2002,Sengar2017}.

\end{itemize}

Despite their different progenitors, all channels lead to systems with low-mass, hydrogen-deficient donors and secularly decreasing mass-transfer rates. Conservation of angular momentum results in long-term orbital expansion, with $P_{\rm orb}$ potentially increasing to $\sim 120$ minutes within a Hubble time \citep{vanHaaften2012b}. As the accretion rate declines, the systems become progressively fainter, providing access to the low-$\dot{M}$ regime of NS accretion in hydrogen-poor environments \citep{Nelemans2010,Heinke2013}.

\subsubsection{UCXBs in Globular Clusters}

Globular clusters are efficient environments for the formation of compact binaries, with LMXBs occurring at rates significantly higher per unit stellar mass than in the Galactic field \citep{Katz1975}. Dynamical interactions such as tidal captures, direct collisions, and exchange encounters \citep{Fabian1975,Hills1976,Verbunt1987,Rasio2000,Ivanova2005} can seed compact NS binaries with donor compositions distinct from those of field systems; however, tidally captured binaries are expected to form with relatively long orbital periods and donors that still retain large convective envelopes, so substantial subsequent orbital shrinkage is required before such systems reach ultracompact periods. From an accretion standpoint,the mass-transfer rate as a function of orbital period should differ little between cluster and field UCXBs when the donor is a genuine white dwarf, since this relation is governed primarily by the white-dwarf equation of state and gravitational-wave losses; larger differences can arise for He-star donors, whose mass--radius relation departs from the degenerate white-dwarf track, though the He-star and white-dwarf tracks generally converge at long orbital periods. A different relative fraction of He-WD versus CO-WD donors between clusters and the field could therefore still shift the persistent/transient boundary between the two populations, although evidence for both donor types exists in both environments \citep{Zurek2009,Heinke2013}.

\subsubsection{Observational Signatures and Accretion Diagnostics}

Given the difficulty of directly measuring very short orbital periods, UCXBs are often identified through accretion-related diagnostics:

\begin{itemize}
\item Spectral Composition: Optical and X-ray spectra typically lack hydrogen features and instead exhibit helium- and metal-rich emission lines, particularly nitrogen and carbon \citep{Nelemans2004,Baglio2014,ArmasPadilla2020,Stoop2021}. This reflects the chemical composition of the transferred material and influences both disc opacity and nuclear burning conditions.

\item Xray-to-Optical Flux Ratio: Because the accretion disc in a UCXB is physically very small, these systems typically show an anomalously high X-ray-to-optical flux ratio compared with LMXBs hosting more extended discs, providing a simple, purely photometric diagnostic of an ultracompact configuration \citep{Pattie2026}.

\item X-ray Spectroscopy: Enhanced oxygen and neon abundances inferred from X-ray spectra, including strong O\,\textsc{viii} Ly$\alpha$ emission, provide evidence for accretion from degenerate donors and chemically evolved discs \citep{Schulz2001,Juett2001,Krauss2007,Madej2010}. Moreover, a very weak or absent Fe~K$\alpha$ emission line has been proposed and observationally tested as a signature of a carbon--oxygen (or O/Ne/Mg) white-dwarf donor, since oxygen efficiently screens the iron line in such discs, whereas He-rich discs retain a strong Fe~K$\alpha$ line \citep{Koliopanos2013,Koliopanos2014}.

\item Thermonuclear Bursts: For systems with helium-rich donors, the absence of hydrogen significantly modifies ignition conditions on the NS surface, leading to helium-powered Type~I X-ray bursts with characteristic durations, recurrence times, and energetics \citep{Cumming2003,Falanga2008,Galloway2020}. By contrast, systems accreting from carbon--oxygen white-dwarf donors are not expected to burst at all, with the possible exception of rare superbursts. UCXBs thus offer a controlled setting to study helium-rich nuclear burning under steady accretion.

\item Accretion Disc Stability: Owing to their small disc radii, UCXBs can remain persistently accreting at luminosities as low as $L_{X,\rm pers}\lesssim10^{36}\,{\rm erg\,s^{-1}}$, consistent with disc instability theory applied to hydrogen-poor compositions \citep{vanParadijs1994,Lasota2001}. The interplay between disc size, chemical composition, and irradiation is central to understanding their long-term variability \citep{intZand2007}.

\end{itemize}

A complete and continuously updated catalogue of known and candidate UCXBs can be found in \citet{ArmasPadilla2023}.

\subsection{Accreting Millisecond Pulsars} 
\label{sec:amxp} 

Millisecond pulsars (MSPs) are NSs with spin periods of $\sim 1$--$10$ ms and 
relatively weak surface magnetic fields ($B \sim 10^{8-9}$ G). Their formation is understood within the framework of the recycling scenario \citep{BisnovatyiKogan1974,SmarrBlandford1976,Alpar82,Radhakrishnan1982}, in which an old NS is spun up through prolonged accretion in a low-mass X-ray binary (LMXB)  \citep[see also, e.g.,][]{Bhattacharya1991}. In this picture, angular momentum is 
transferred from the accretion disc to the NS via magnetospheric coupling, 
progressively increasing the stellar spin frequency while sustained accretion 
reduces the effective large-scale dipolar magnetic field. Over long timescales, 
accretion torques can spin the NS up to hundreds of rotations per second, 
producing an MSP.

The spin evolution of the NS is governed by the interaction between the disc and 
the stellar magnetosphere. The accretion disc is truncated at the magnetospheric 
radius,
\begin{equation}
r_{\rm m} \simeq \left( \frac{\mu^4}{2GM\dot{M}^2} \right)^{1/7},
\end{equation}
where $\mu$ is the magnetic dipole moment, $M$ the NS mass, and $\dot{M}$ the 
mass-accretion rate. The torque exerted on the star depends critically on the 
relative position of $r_{\rm m}$ and the corotation radius,
\begin{equation}
r_{\rm co} = \left( \frac{GM}{\Omega^2} \right)^{1/3},
\end{equation}
where $\Omega$ is the NS spin angular velocity.

When $r_{\rm m} < r_{\rm co}$, matter from the disc can accrete onto the stellar 
surface, transferring positive angular momentum and spinning up the NS. As the 
star accelerates, the corotation radius decreases, and the system approaches a 
state of spin equilibrium defined approximately by $r_{\rm m} \sim r_{\rm co}$. 
The equilibrium spin frequency therefore depends on both $\dot{M}$ and the 
magnetic field strength, linking the final MSP spin distribution directly to 
long-term accretion physics.

If instead $r_{\rm m} > r_{\rm co}$, centrifugal forces inhibit accretion and 
the system may enter the so-called \textit{propeller regime} 
\citep{Illarionov1975}, in which matter is partially or entirely expelled from 
the system. At even lower accretion rates, the magnetosphere may expand beyond 
the light-cylinder radius, allowing the activation of the rotation-powered radio 
pulsar mechanism. The transitions between these regimes are therefore controlled 
by variations in $\dot{M}$ and represent different modes of disc--magnetosphere 
coupling.

For many years, direct observational confirmation of accretion-driven spin-up 
remained elusive, as coherent millisecond pulsations were not detected in most 
LMXBs. Indeed, the majority of persistently accreting systems --- including 
luminous sources such as Sco~X-1 --- show no detectable coherent pulsations, 
and understanding why remains an open issue \citep{Patruno2021}. Within the 
context of accretion physics, several explanations have been proposed. If the 
magnetic field is sufficiently weak, the magnetospheric radius may lie very close 
to the NS surface, preventing efficient channelling of the accretion flow onto 
the magnetic poles. Alternatively, prolonged accretion may partially bury the 
magnetic field beneath freshly accreted material, reducing the observable 
pulsation amplitude \citep{Romani1990,Cumming2001}. Scattering in optically 
thick accretion flows or boundary-layer emission may further dilute pulsations 
\citep{Brainerd1987}.

The detectability of pulsations is also expected to depend on the accretion rate. 
Most LMXBs are transient systems, undergoing large luminosity variations and 
spectral state transitions. During low-luminosity, hard states, when $\dot{M}$ 
decreases and the inner disc may recede, the magnetosphere can expand, 
potentially enhancing magnetic channelling and making millisecond pulsations 
detectable \citep[e.g.,][]{Gogus2007}. The appearance or disappearance of 
pulsations thus provides insight into the geometry of the inner accretion flow.

An important milestone came in 1998 with the discovery of coherent millisecond X-ray pulsations in the transient LMXB SAX J1808.4--3658 \citep{Wijnands98} --- the first source to show persistent, coherent pulsations directly revealing the NS spin, complementing the burst oscillations already discovered two years earlier \citep{Strohmayer1996}. This system established the class of accreting millisecond X-ray pulsars (AMXPs; 
see \citealt{disalvo2022,Patruno2021,Campana2018} for reviews), demonstrating 
that NSs in LMXBs can reach millisecond spin periods while actively accreting. 
Since then, 26 AMXPs have been identified (Table~\ref{tab:amxps}), providing 
laboratories for studying magnetically channelled accretion, torque variability, 
and spin equilibrium at low accretion rates. Three of these are transitional 
millisecond pulsars (tMSPs; Sect.~\ref{tmsp_sec}), which alternate between 
accretion-powered and rotation-powered states, providing the clearest 
observational evidence that the same NS can function as both an accreting LMXB 
and a radio MSP.

A few sources in Table~\ref{tab:amxps} are intermittent AMXPs, in which pulsations are detected only sporadically. In Aql X-1, coherent pulsations were detected only during a $\sim$150~s episode in a single outburst \citep{Casella_2008ApJ}, while in SAX J1748.9$-$2021 they appeared intermittently during outbursts \citep{Altamirano2008}, and in HETE J1900.1$-$2455 they faded within months of the outburst onset while accretion continued \citep{Galloway2007}. The mechanisms usually invoked to explain the general non-detection of pulsations in most NS LMXBs do not obviously account for this behaviour, which remains only partly understood.

NSs are typically born with high magnetic fields ($B\sim10^{12\text{--}13}$~G) and spin periods of tens to hundreds of milliseconds. In isolation, they spin down due to magnetic dipole radiation and particle winds, evolving toward longer periods and eventually entering the so-called pulsar graveyard. In contrast, sustained disc accretion in LMXBs spins the NS back up, transferring angular momentum to the star and reshaping its magnetic-field configuration. The location of MSPs in the $P$--$\dot{P}$ diagram therefore reflects the cumulative impact of long-term accretion torques and magnetic-field evolution \citep{Tauris2012}.

\begin{figure}
\centering
\includegraphics[width=1\linewidth]{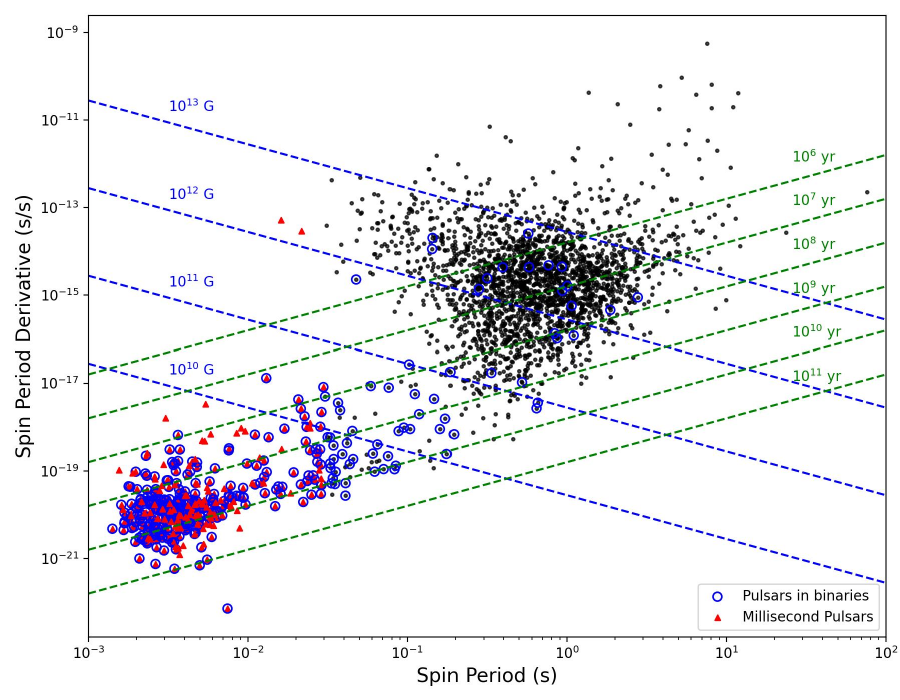}
\caption{$P$--$\dot{P}$ diagram of rotation-powered pulsars. Blue circles mark 
pulsars in binary systems. Red triangles mark millisecond pulsars (here defined 
as pulsars with a spin period $< 30$ ms). Lines of constant magnetic field 
\citep[evaluated as $B\sim 6\times 10^{19}(P\dot{P})^{1/2}$ G;][]{disalvo2022} 
and spin-down age ($P/2\dot{P}$~s) are plotted with blue and green dashed lines, 
respectively.}
\label{fig:pulsars}
\end{figure}

\begin{sidewaystable}[]
    \centering
    \resizebox{\textheight}{!}{%
    \begin{tabular}{l|c|c|c|c|l|>{\raggedright\arraybackslash}p{3.5cm}}
    \hline    \hline
    \multicolumn{7}{c}{Accreting Millisecond X-ray Pulsars} \\
    \hline
    Source & $\nu_{\rm S}$ (Hz) & $P_{\rm orb}$ (days) & $f_{\rm x}\, (M_{\odot})$ & $M_{\mathrm{c,min}}\, (M_{\odot})$ & Companion type & Ref. \\
    \hline
    IGR J00291+5934 & 599 & 0.1025 & $2.8\times 10^{-5}$ & 0.039 & BD & \citet{Galloway2005} \\
    MAXI J0911$-$655 & 340 & 0.031 & $6.2\times10^{-6}$ & 0.024 & He WD (?) & \citet{sanna2017} \\
    XTE J0929$-$314 & 185 & 0.0304 & $2.9\times10^{-7}$ & 0.0083 & C/O WD & \citet{Galloway2002,Giles2005} \\
    XTE J1751$-$305 & 435 & 0.0296 & $1.3\times10^{-6}$ & 0.014 & He WD & \citet{Markwardt_2002,Davanzo2009} \\
    SRGA J144459.2$-$604207 & 448 & 0.2176 & $6.3\times10^{-3}$ & 0.25 & MS? & \citet{Ng2024,Molkov2024} \\
    IGR J16597$-$3704 & 105 & 0.0321 & $1.2\times10^{-7}$ & 0.006 & He WD & \citet{Sanna2018_597} \\
    IGR J17062$-$6143 & 164 & 0.0263 & $9.1\times10^{-8}$ & 0.006 & He WD? & \citet{Strohmayer2017} \\
    IGR J17379$-$3747 & 468 & 0.0783 & $8\times10^{-5}$ & 0.056 & BD & \citet{Sanna2018_1737} \\
    IGR J17494$-$3030 & 376 & 1.25 & $1.4\times10^{-6}$ & 0.14 & He or C/O WD & \citet{Ng2021} \\
    SAX J1748.9$-$2021$^{\dagger}$ & 442 & 0.365 & $4.8\times10^{-4}$ & 0.1 & MS & \citet{Altamirano2008,Cadelano2017} \\
    NGC 6440 X-2 & 206 & 0.04 & $1.6\times10^{-7}$ & 0.0067 & He WD & \citet{Altamirano2010} \\
    SWIFT J1749.4$-$2807 & 518 & 0.369 & $5.5\times10^{-2}$ & 0.59 & MS & \citet{altamirano2011,Davanzo2011} \\
    IGR J17498$-$2921 & 401 & 0.16 & $2.0\times10^{-3}$ & 0.17 & MS & \citet{Papitto2011} \\
    IGR J17511$-$3057 & 245 & 0.1446 & $1.1\times10^{-3}$ & 0.13 & MS & \citet{Papitto2010} \\
    Swift J1756.9$-$2508 & 182 & 0.0379 & $1.6\times10^{-7}$ & 0.007 & He WD & \citet{Krimm2007} \\
    IGR J17591$-$2342 & 527 & 0.3667 & $1.5\times10^{-2}$ & 0.37 & MS & \citet{Sanna_2018AA_IGRJ17591-2342} \\
    XTE J1807$-$294 & 190 & 0.0279 & $1.5\times10^{-7}$ & 0.0066 & C/O WD & \citet{Campana_2003ApJ,Davanzo2009} \\
    SAX~J1808.4$-$3658 & 401 & 0.0837 & $3.8\times10^{-5}$ & 0.043 & BD & \citet{Wijnands98,wang2013} \\
    XTE J1814$-$338 & 314 & 0.178 & $2.0\times10^{-3}$ & 0.17 & MS & \citet{Markwardt2003,Wang2017} \\
    MAXI J1816$-$195 & 529 & 0.20125 & $4.8\times10^{-3}$ & 0.10 & MS? & \citet{Bult_2022ApJ} \\
    HETE J1900.1$-$2455$^{\dagger}$ & 377 & 0.0579 & $2.0\times10^{-6}$ & 0.016 & BD & \citet{Kaaret_2006ApJ,Elebert_2008MNRAS} \\
    Aql X-1$^{\dagger}$ & 550 & 0.7896 & $1.4\times10^{-2}$ & 0.56 & MS & \citet{Casella_2008ApJ,MataSanchez_2017MNRAS} \\
    MAXI J1957+032 & 314 & 0.04208 & $1.6\times10^{-6}$ & 0.017 & MS & \citet{Sanna_2022MNRAS} \\
    \hline
    \multicolumn{7}{c}{Transitional Millisecond Pulsars}\\
    \hline
    IGR J18245$-$2452$^{*}$ & 254 & 0.4596 & $2.3\times10^{-3}$ & 0.17 & MS & \cite{Papitto2013} \\
    PSR J1023$+$0038 & 592 & 0.1979 & $1.1\times10^{-3}$ & 0.14 & MS & \citet{Archibald2009} \\
    XSS J12270$-$4859 & 593 & 0.28792 & $3.9\times10^{-3}$ & 0.15 & MS & \citet{Bassa2014,deMartino2015} \\
    \hline
    \end{tabular}}    
\caption{List of accreting millisecond X-ray pulsars and transitional millisecond pulsars. Here $\nu_{\rm S}$ is the spin frequency, $P_{\mathrm{orb}}$ is the orbital period, $f_{\rm x}$ is the X-ray mass function, and $M_{\mathrm{c,min}}$ is the minimum companion mass, calculated for a system inclination $i=90^{\circ}$ and assuming a NS mass of 1.4~M$_\odot$. The types of companions are as follows: WD = White Dwarf (He WD = Helium WD, C/O WD = Carbon/Oxygen WD), BD = Brown Dwarf, MS = Main Sequence star. $^{*}$This is the only transitional millisecond pulsar that has shown accretion-related X-ray millisecond pulsations \citep{Papitto2013} and can therefore be regarded as an accreting millisecond X-ray pulsar. Updated from \citet{disalvo2022}. $^{\dagger}$Intermittent AMXPs (see Sect.\ref{sec:amxp}).}
\label{tab:amxps}
\end{sidewaystable}

\subsubsection{Transitional millisecond pulsars}
\label{tmsp_sec}

To date, three confirmed tMSPs are known: PSR\,J1023+0038, XSS\,J12270$-$4859, 
and IGR\,J18245$-$2452 \citep[see][and references therein]{Campana2018,Papitto2022}, 
along with a few additional candidates. PSR\,J1023+0038 was the first source 
identified as a potential link between LMXBs and radio MSPs: initially detected 
in the radio band \citep{Bond2002}, it later showed optical signs of an accretion 
disc \citep{Szkody2003}, and in 2007 a 1.69~ms radio pulsar was discovered as its compact object \citep{Archibald2009}. The system underwent a state change to an optical accretion-disc dominated state in June 2013 \citep{Stappers2014}, accompanied by an increase in $L_{\rm X}$ to $\sim2.5\times10^{33}$~erg\,s$^{-1}$ \citep{Patruno2014}, but has never displayed a full (e.g. $L_{\rm X}>10^{35}$~erg\,s$^{-1}$) accretion outburst. XSS J12270--4859 followed a similar pattern: radio, optical and X-ray observations indicated a switch to a rotation-powered MSP phase since the end of 2012 \citep{Bassa2014,Bogdanov2014,deMartino2015}, confirmed by the detection of 1.69~ms radio pulsations \citep{Roy2015}, though no full X-ray outburst was observed in either system. The clearest observational evidence for the LMXB--MSP link came from IGR J18245$-$2452, the only confirmed tMSP to have undergone a full (peak 2--10 keV $L_{\rm X}\sim6\times10^{36}$~erg\,s$^{-1}$) accretion outburst \citep{Papitto2013}, followed weeks later by the reactivation of the radio pulsations.
These systems occupy a critical region of parameter space in which the 
mass-accretion rate is sufficiently low that the interaction between the 
accretion flow and the NS magnetosphere becomes finely balanced. As a result, 
modest variations in $\dot{M}$ can shift the system between rotation-powered and 
accretion-powered regimes.

Among the confirmed systems, IGR~J18245$-$2452 is the only source observed to 
undergo a full transition from a rotation-powered radio pulsar state to an 
accretion-powered X-ray outburst \citep{Papitto2013}. The other two systems have 
been observed to enter a distinct sub-luminous X-ray state, characterized by 
$L_{\rm X} \sim 10^{33}$--$10^{34}$ erg s$^{-1}$, during which radio pulsations 
disappear and a disc is present. This state was initially interpreted as 
accretion-powered, but its phenomenology indicates that the system resides in a 
regime where the magnetospheric radius is comparable to the corotation radius. In 
this configuration, the accretion flow is neither fully suppressed nor steadily 
reaching the stellar surface, leading to rapid mode switching and strong 
variability.

PSR~J1023+0038, in particular, has remained in this sub-luminous disc state 
since 2013 \citep{Stappers2014}, providing a unique laboratory for studying the 
interplay between accretion inflow and magnetospheric pressure. In this regime, 
the system appears to alternate between episodes in which matter penetrates inside 
the corotation radius and episodes in which centrifugal inhibition or outflows 
dominate, possibly operating in a weak propeller configuration. The luminosity, 
variability, and multi-wavelength behaviour of this state therefore reflect the 
system hovering near the boundary defined by $r_{\rm m} \sim r_{\rm co}$, where 
small fluctuations in $\dot{M}$ can produce large changes in the accretion 
geometry.

While XSS~J12270$-$4859 and IGR~J18245$-$2452 are currently in their radio 
pulsar states \citep{Campana2018,Papitto2022}, PSR~J1023+0038 continues to offer 
an ongoing view of this delicately balanced regime. The extensive observational 
efforts devoted to this source have revealed that transitional MSPs are not simply 
systems switching between two distinct states, but rather NSs operating in a 
narrow accretion-rate window where disc--magnetosphere coupling, propeller action, 
and rotation-powered emission coexist and compete. In the following, we summarize 
the most recent results on PSR~J1023+0038, which continues to challenge and 
refine our understanding of how accretion and magnetospheric processes regulate 
the transition between accretion-powered and rotation-powered NSs.

\vspace{5pt}
\noindent
\textbf{The sub-luminous disc state of PSR J1023+0038}
\vspace{5pt}

While in the sub-luminous disc state, PSR\,J1023+0038 exhibits frequent mode switchings between three different X-ray intensity levels \citep[e.g.][]{Papitto2013, Linares2014, Bogdanov2015}: a \textit{high} luminosity mode (for $\sim70\%$ of the time), where coherent X-ray pulsations are detected \citep{Archibald2015}; a \textit{low} luminosity mode ($\sim30\%$ of the time, $\sim 7$ times dimmer than the high mode), with no detection of X-ray pulsations; and a \textit{flaring} mode ($\sim$ 2$\%$ of the time), during which the X-ray luminosity increases by around an order of magnitude compared to the high mode. The switches between the high and the low luminosity modes happen on a timescale of $\lesssim10$~s, and are visualized in the X-ray light curve as rectangular-shaped dips. Similar switches have been observed at UV and optical frequencies in the case of PSR J1023+0038 \citep{Shahbaz2015,MiravalZanon2022,Jaodand2021}. A bright emission in the radio band with a flat spectrum was detected by \citet{Deller2015} during a long observation encompassing both high and low modes, and was interpreted as due to the emission of a compact, self-absorbed jet.  Later, radio flares were observed in correspondence with transitions to the X-ray low mode \citep{Bogdanov2018}.

This strong anti-correlation between the radio and the X-ray light curves was interpreted as due to the occurrence of additional short-lived mass ejection processes occurring close to the NS.
Optical and UV pulsations at the pulsar spin period were observed, making 
PSR~J1023+0038 the first optical/UV MSP ever detected \citep{Ambrosino2017,
Papitto2019,MiravalZanon2022}, implying that particle 
acceleration by the rotation-powered pulsar wind persists even in the presence of 
accreting matter. Optical pulsations were subsequently detected also in 
SAX~J1808.4$-$3658 during outburst \citep{Ambrosino2021}, suggesting that this 
mechanism may operate more broadly among AMXPs, though its interplay with 
accretion remains poorly understood. It was suggested that PSR~J1023+0038 might 
be active as a rotation-powered pulsar emitting a magnetized wind that interacts 
with the accreting matter close to the light cylinder radius. At this boundary, electrons are accelerated producing optical, UV and X-ray 
pulsations through synchrotron radiation. This picture is supported by several 
independent lines of evidence: the detection of polarized X-rays from the 
boundary during high modes \citep[][see Sect.~\ref{sec:ixpe}]{BaglioIXPE}, 
the measurement of X-ray time lags consistent with synchrotron emission at the 
pulsar wind boundary \citep{Illiano2023}, and estimates of the spin-down power 
indicating that the pulsar wind carries sufficient energy to power the observed 
emission \citep{Burtovoi2020}. This interpretation also accounts for the 
emission in the \textit{flaring} mode, which may arise from transient increases 
in the effective area of the disc exposed to the wind, temporarily enhancing the 
interaction and emission \citep{Veledina2019}. In the \textit{low} mode, 
\cite{Papitto2019} suggested that the boundary is moved at larger radii by the 
pulsar wind. In contrast, \cite{Veledina2019} hypothesized that the accretion 
disc may cross into the light cylinder during the low mode. When this occurs, the 
pulsar wind no longer dissipates energy via direct collision with the disc, and 
the system likely transitions into a propeller regime. A dramatic variability in 
the NIR emission was also detected, including strong and rapid flares 
\citep{Baglio2019}. The NIR emission was found to lag the optical emission by 
about 10~s, interpreted as reprocessing of the optical emission by matter ejected 
by the active rotation-powered pulsar \citep{Baglio2019}.

Recently, very short duration ($\sim 15$s) mm flares were detected thanks to ALMA observations  \citep{Baglio2023}. Such flares were observed in correspondence to the start of X-ray low-mode episodes, possibly anticipating the longer radio flares, and were interpreted as the signature of the launching of optically thin discrete ejecta. 
A detailed modelling of the spectral energy distributions (SEDs) extracted separately in the high and low modes using the data of the same campaign allowed to put forward a scenario where the high-to-low mode switch was interpreted as the evolution of the innermost accretion disc region that drives emission of discrete ejecta on top of a compact jet, and the low-to-high mode switch as due to the re-enshrouding of the pulsar \citep{Baglio2023}. This scenario is mostly in agreement with the mini-pulsar wind nebula scenario reported by \citet{Papitto2019}, as the ejection of the innermost accretion flow at the transition to the low mode supports a shift of the boundary region outward, towards the outer disc.

The flaring mode has been poorly investigated. The most recent studies on PSR J1023+0038 \citep{Baglio1023flaring} showed that X-ray flares seem to be accompanied by a UV and optical brightening and to be aligned with intense and long radio flares (a 1-hour duration radio flare peaking at 1.2 mJy at 6 GHz in correspondence to a similar duration X-ray flaring mode was reported in \cite{Baglio1023flaring}). A notable drop in the optical polarization during the flare, from around polarization degree (PD) of 1.4\% to 0.5\%, pointed to the appearance of an unpolarized emission component, likely linked to structural changes in the accretion disc\, consistent with the substantial increase in the amplitude of optical pulsations reported during the flare by \cite{Karpov2026}.

\vspace{5pt}
\noindent
\textbf{Transitional millisecond pulsar candidates}
\vspace{5pt}

One of the most effective strategies for identifying new candidate tMSPs has 
involved targeting systems that exhibit emission characteristics resembling those 
of the distinctive subluminous disc state observed in all known tMSPs. This state 
is characterized by a rich and distinctive phenomenology. In the X-ray band, the 
spectrum is well described by a power law with photon index $\Gamma \sim 1.6$--$1.7$ 
\citep{Campana2016}, and the flux switches rapidly between a \textit{high} mode and a 
\textit{low} mode on timescales of tens of seconds, often accompanied by erratic 
flaring episodes across the X-ray, UV, optical, and infrared bands. The presence 
of an accretion disc is confirmed by double-peaked optical emission lines, 
a hallmark of Keplerian rotation in the disc \citep[e.g.][]{CotiZelati2014,DeMartino2014}. Perhaps most 
surprisingly, tMSPs in the subluminous disc state are strong gamma-ray emitters, 
with emission extending to GeV energies detected by the \textit{Fermi} satellite 
\citep{Stappers2014, Takata2014, Torres2022}. This is remarkable for two reasons: low-mass X-ray binaries are not normally GeV gamma-ray sources, and the gamma-ray luminosity of tMSPs in this state exceeds that of typical rotation-powered radio pulsars, implying that particle 
acceleration proceeds with extraordinary efficiency even in the presence of 
accreting material. Strong radio continuum emission is also observed, and in 
PSR~J1023+0038 the radio and X-ray light curves show a striking anti-correlation, 
interpreted as evidence for short-lived mass ejection events occurring close to 
the NS \citep{Bogdanov2018}. The combination of these properties defines a recognizable 
observational fingerprint that has guided the search for new tMSP candidates.

Among the most recently discovered tMSP candidates we can find systems that have been the subject of focused multi-wavelength campaigns in the past few years. Since the properties and discovery history of several candidates have already been extensively reviewed elsewhere \citep[e.g.,][]{Papitto2022}, here we limit our discussion to summarizing the latest developments and newly proposed systems that contribute to expanding and refining the class of tMSP candidates.

\begin{itemize}
\item RXS~J154439.4$-$112820: the first tMSP candidate to be identified 
\citep{Bogdanov2015_1544}, this source was found within the positional uncertainty 
of 3FGL~J1544.6$-$1125 and displays several hallmarks of a tMSP in a 
subluminous disc state. Its optical spectrum reveals strong H and He emission 
lines, consistent with an active accretion disc. The X-ray light curve exhibits 
rapid mode switching on $\sim$10-second timescales, with flux levels differing 
by about a factor of 10, while the spectrum follows a power law with photon 
index $\Gamma \simeq 1.7$ and minimal spectral variability --- typical of tMSPs 
\citep{Bogdanov2015_1544,Bogdanov2016}. Long-term optical monitoring shows variability 
of $\sim$0.5 mag, with occasional flares reaching up to 1.5 mag, closely 
resembling the flaring behaviour seen in known tMSPs. Variable, flat-spectrum 
radio emission has also been detected, placing the source in the same region of 
the radio/X-ray luminosity plane as other tMSPs \citep{Jaodand2021_1544}. 
Time-resolved optical spectroscopy revealed a 5.8-hour orbital period and a very 
low inclination ($5^\circ$--$8^\circ$), suggesting a nearly face-on geometry 
\citep{Britt2017}. Using a distance estimate of 2.8 kpc based on Gaia eDR3 and 
a Galactic prior, the source's X-ray and $\gamma$-ray luminosities are 
$\sim8.1\times10^{33}$\,erg\,s$^{-1}$ and $1.3\times10^{34}$\,erg\,s$^{-1}$, 
respectively --- both consistent with other systems in the tMSP subluminous disc 
state. Strikingly, a recent multi-wavelength campaign revealed a radio/X-ray 
anti-correlation \citep{Gusinskaia2025}, making RXS~J154439.4$-$112820 the only 
source besides PSR~J1023+0038 known to exhibit this behaviour, and further 
strengthening its identification as a tMSP candidate. Clear X-ray mode switching 
and correlated optical variability pointing to a shared origin of the emission at 
the boundary between the pulsar wind and the accretion flow were also reported \citep{illiano2025}.

    \item CXOU J110926.4‑650224, currently observed in a low-luminosity accretion disc state. Evidence for its classification includes the detection of an accretion disc, distinct transitions between X-ray intensity modes, and episodic flaring activity in the X-ray band. Additionally, the system exhibits variable radio continuum emission, with some radio flares showing a temporal connection to those seen in X-rays \citep{cotizelati2019,cotizelati2021}. However, in-depth investigation of J1109’s fast variability across wavelengths is complicated by observational challenges: the source is intrinsically faint in X-rays, optical, and NIR bands, primarily due to its considerable distance ($\sim4$ kpc), and its location in a densely populated stellar field further hampers efforts to resolve its short-term multi-band behaviour. Recently, \cite{cotizelati2024} showed strong optical flickering, redder dips, and bluer flares, with each dip corresponding to an X-ray low mode. This result is in agreement with the scenario put together by \citet{Baglio2023} for the high/low mode transition, and indicates a co-spatial origin for much of the emission. 
    \item 4FGL J0639.1$-$8009, discovered in March 2025 by \citet{Kyer2025}, it lies within the error ellipse of an unassociated \emph{Fermi}-LAT gamma-ray source. It shows strong, variable X-ray emission characterized by a power-law spectrum with a photon index of $\Gamma \sim 1.7$--$1.9$, which is typical for tMSPs in the subluminous disc state. Optical observations reveal rapid, high-amplitude variability unrelated to orbital motion, along with persistent accretion disc-like spectral features. These properties are inconsistent with those of accreting white dwarfs and support the identification of the system as a tMSP candidate.
    \item 4FGL J1824.2$+$1231, also associated with an unclassified \emph{Fermi}-LAT gamma-ray source \citep{Kyer2025}, it exhibits similar behaviour to 4FGL J0639.1$-$8009. Its X-ray emission is variable and follows a power-law distribution with photon indices between $\Gamma \sim 1.7$ and $1.9$. The optical emission is highly variable on short timescales and does not show signs of modulation from orbital dynamics. Spectroscopy reveals accretion disc features. Like the other candidate, these combined characteristics are aligned with the known subluminous disc state seen in confirmed tMSPs, suggesting that 4FGL J1824.2$+$1231 is also a strong tMSP candidate.
    \item 4FGL J0427.8-6704 \citep{Strader2016}, so far only been observed in the subluminous disc state. The system exhibits deep eclipses across optical, X-ray, and gamma-ray wavelengths, implying a high orbital inclination ($i\approx84^\circ$). Its distance, estimated from Gaia DR3 and optical modelling, is approximately 2.5 kpc. The source displays flaring X-ray behaviour without clear mode switching, and its soft X-ray luminosity spans $(0.5$--$20)\times10^{33}$ erg s$^{-1}$---comparable to PSR J1023+0038. The X-ray spectrum is best described by a partially absorbed power-law with photon indices $\Gamma\sim1.2$--$1.8$, and high covering fractions, suggesting obscuration by the accretion disc seen edge-on \citep{Kennedy2020}. The system's short orbital period (8.8 h), component masses (NS $\sim1.4\,M_\odot$, companion $\sim0.3\,M_\odot$), and persistent gamma-ray eclipses indicate that high-energy emission originates very close to the compact object. Multiwavelength observations, including ATCA and ALMA radio data, reveal a flat radio spectrum with average flux densities of $\sim$0.3--0.4 mJy and short-timescale flaring up to 2~mJy. The radio emission lacks eclipses, supporting a compact jet origin. Modelling of the spectral energy distribution performed by \cite{Koljonen2025} revealed a jet break frequency around $10^{11}$ Hz, from which they inferred a magnetic field strength at the base of the jet of $B_0 \sim$100~ G, and a base radius of $R_0 \sim 10^{10}$~cm. No eclipses were observed in the mm band, suggesting a jet launch zone located at $z_0 > 7 \times 10^{10}$~cm from the NS.

\end{itemize}


\section{Variability of NS XRBs}\label{sec3}

\subsection{Short timescale variability}\label{sec:variab}

\begin{figure}
\centering
\includegraphics[width=0.5\linewidth]{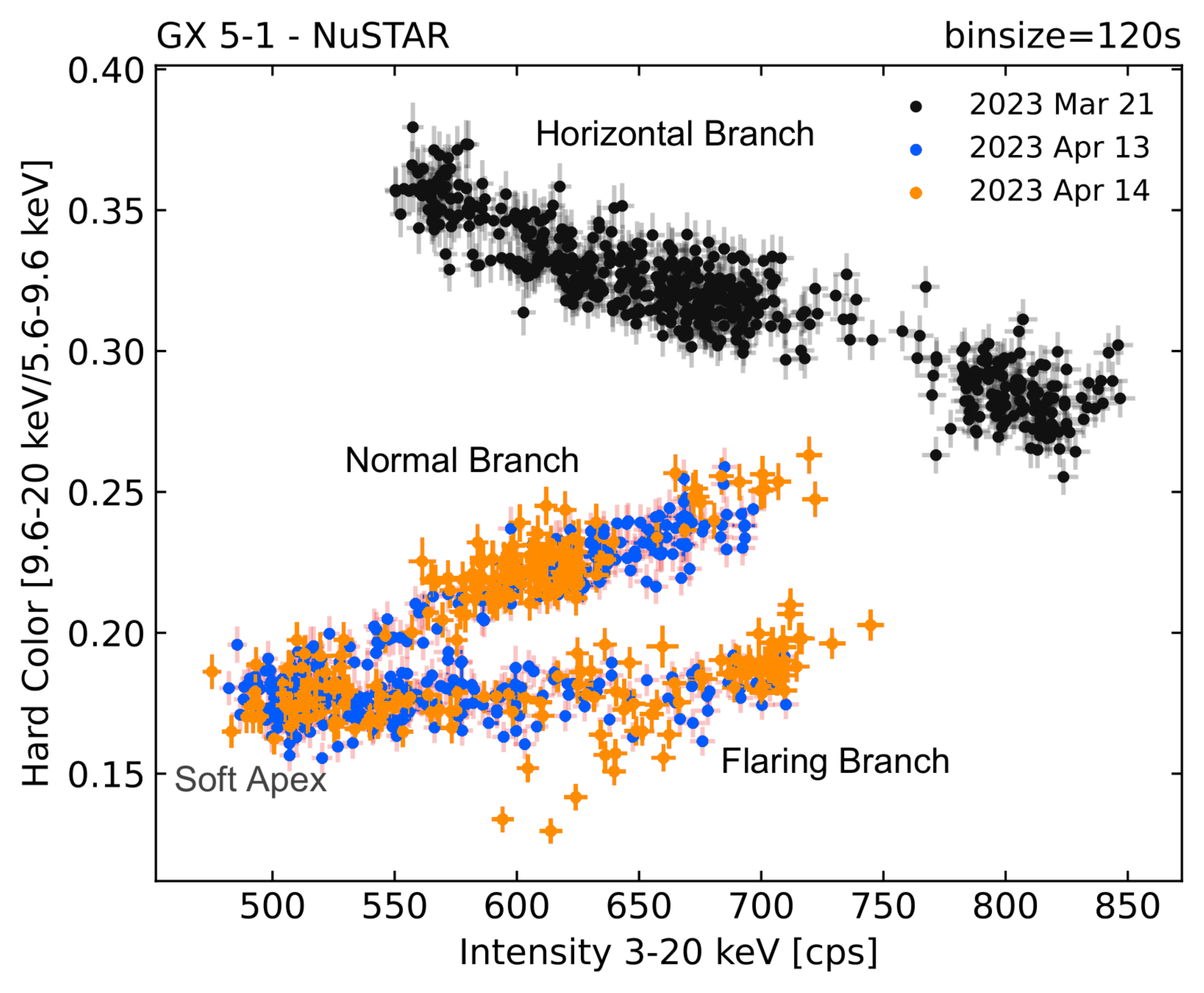}
\includegraphics[width=0.45\linewidth]{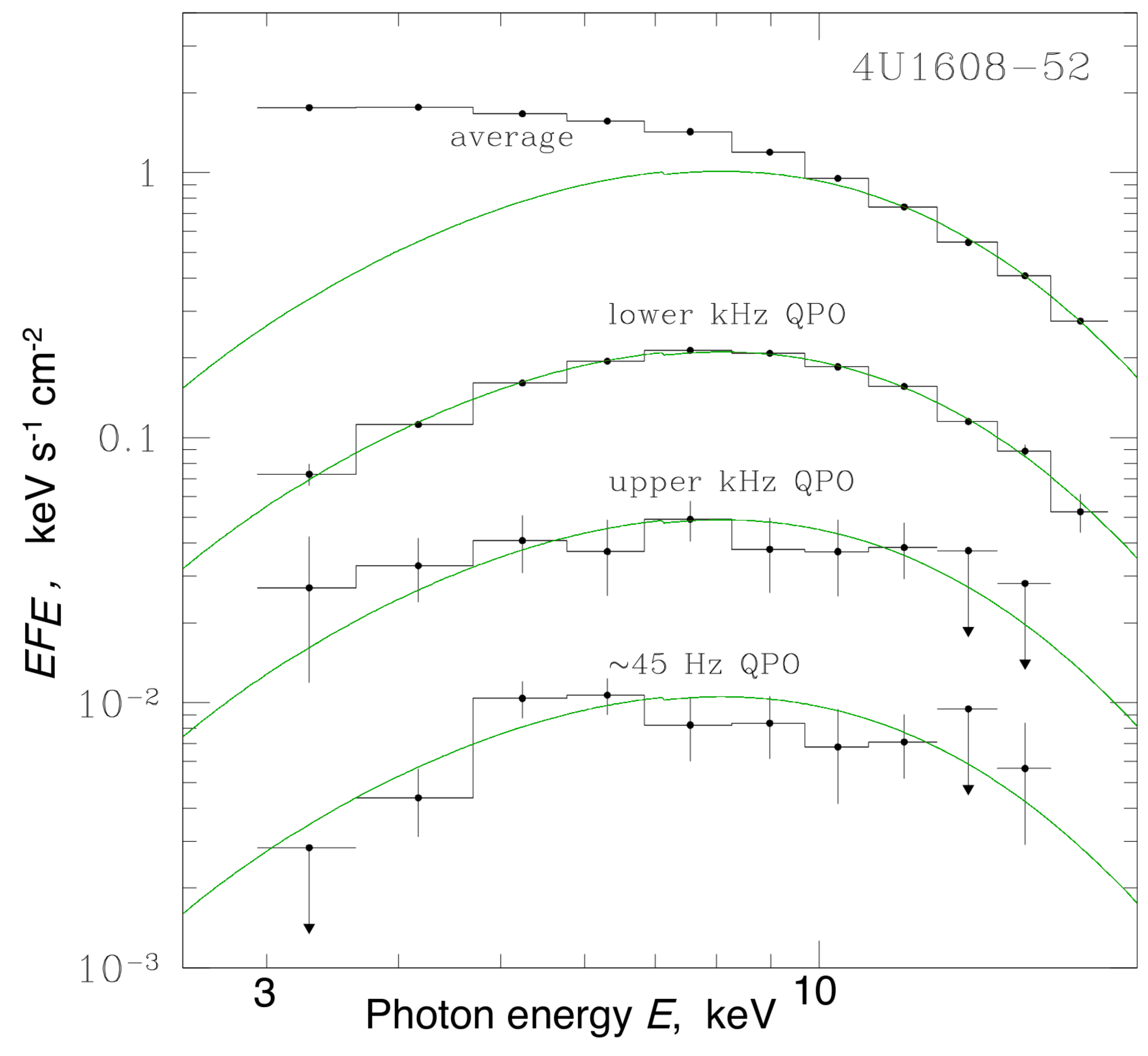}
\caption{Left: Z-track of GX~5$-$1 at the hardness-intensity diagram (adapted from \citealt{Fabiani2024}).  Right: Fourier-frequency resolved spectra of various QPOs in 4U~1608$-$52 (adapted from \citealt{Gilfanov2003}).}
\label{fig:HID_qpos}
\end{figure}

Short-time-scale (milliseconds to hours) X-ray variability in NS-LMXBs traces the coupling between the inner accretion flow, the boundary layer, and the stellar surface. The X-ray power density spectra (PDS) comprise broad-band noise components and quasi-periodic oscillations (QPOs), whose characteristic frequencies and amplitudes evolve in time and are correlated with the accretion state and accretion rate. In atoll sources, the band-limited noise dominates in the ``island'' state (which has a relatively hard spectrum) and weakens as variability shifts to higher frequencies in the ``banana'' state (which is an intermediate to soft state); Z-sources (Fig.\,\ref{fig:HID_qpos}, left) trace distinct timing signatures along the horizontal, normal, and flaring branches \citep{Hasinger1989, vanderKlis2006, Done2007}, being more variable along the horizontal branch (HB, the harder state), and progressively less variable along the normal (NB) and then flaring branches (FB), spectrally intermediate and soft, respectively. These trends are commonly interpreted as changes in the disc truncation radius and/or in  the coronal/boundary-layer geometry and/or energetics.

Kilohertz QPOs ($\sim300$--$1200$~Hz) are commonly observed in the PDS from NS--LMXBs, and are among the most diagnostic short-time-scale features, often appearing as paired peaks whose frequencies track the motion of matter very close to the NS surface. Frequency separations comparable to the spin or half-spin motivate beat-frequency and relativistic-precession interpretations, linking the signals to strong-gravity dynamics at small radii \citep{Miller1998, Stella1998, Stella1999, Wijnands2003}.
The Fourier-frequency resolved spectroscopy univocally demonstrates that QPOs are associated with the harder Comptonized emission (\citealt{Gilfanov2003,Revnivtsev2006}; see Fig.\,\ref{fig:HID_qpos}, right). 
The spectrum of this variable component is virtually independent of the accretion rate \citep{Revnivtsev2013} and is consistent with the emission from  a boundary layer or a spreading layer at the NS surface \citep{Suleimanov2006}.  
The observed maximum colour temperature of $kT\sim2.5$~keV is very close to the maximal temperature observed from the photospheric radius expansion X-ray bursts (see below), which is limited  by the local Eddington flux at the NS surface. 
Luminosity variations can then be  associated with changes in the size of the emission region. 
The evolution of coherence and amplitude with state constrains the radiative coupling of the corona and boundary layer \citep{Barret2005, Mendez2006}, while the energy-dependent rms and lags of the kHz QPOs further isolate the variable Comptonized component \citep{Peirano2021, Peirano2022}.

Fourier-resolved spectroscopy, covariance spectra, and inter-band time and phase lags indicate a combination of inward-propagating accretion-rate fluctuations and rapid reprocessing. Hard lags (i.e. where the higher energy emission lags the lower energy one) at low Fourier frequencies are consistent with radially stratified Comptonization and propagating fluctuations (similar to the BH sources, \citealt{Kotov2001}), while soft lags at higher frequencies point to prompt reflection/reprocessing in the disc and from the stellar surface \citep{Arevalo2006, Uttley2014}. Taken together -- broad-band noise, low-/high-frequency QPOs, and burst oscillations (see Sect.~\ref{sec:bursts}) -- these timing diagnostics complement spectroscopy to track geometry, measure spin, and place bounds on the NS mass–radius relation using time-domain constraints \citep{Psaltis1999, vanderKlis2006}.

\subsection{Type-I bursts}
\label{sec:bursts}

First identified as a distinct phenomenon by \citet{Belian1972}, \citet{Babushkina1975} and \citet{Grindlay1976}, Type-I X-ray bursts are a characteristic manifestation of accretion onto NSs, arising from thermonuclear instabilities in the freshly accreted surface layers. They are observed as short (seconds to minutes) flashes of X-rays displaying a fast rise and slower decay, accompanied by spectral softening consistent with the cooling of a blackbody-like emitter (see reviews by \citealt{Lewin1993}, \citealt{Strohmayer2006} and, more recently, \citealt{Degenaar2018}). In some cases, the luminosity approaches the Eddington limit, driving photospheric radius expansion (PRE) episodes that can be used to infer distances \citep{Kuulkers2003} and constrain NS radii \citep{Lewin1993}.
The recurrence behaviour, energetics, and spectral evolution of the bursts depend on the local accretion rate, the composition of the accreted fuel, and the thermal structure of the envelope. Type-I bursts therefore provide a powerful diagnostic of nuclear burning processes under extreme conditions and remain an unambiguous signature of accretion onto a NS.

In addition to probing nuclear physics, Type-I bursts also offer a unique window into the coupling between surface burning and the accretion flow. Simultaneous \textit{INTEGRAL} and ATCA observations of 4U~1728$-$34 and 4U~1636$-$536 revealed minutes-long radio flares following each burst, with frequency-dependent delays consistent with propagation along a compact jet \citep{Russell2024}. From these lags, the jet velocity was measured as $v = 0.38^{+0.11}_{-0.08}c$, providing the first direct estimate for a NS compact jet and showing it to be significantly slower than those in BH systems. These results suggest that the jet responds to burst-induced changes in the inner accretion flow, possibly mediated by variations in the magnetic field configuration anchored in the disc. 

At the same time, bursts can directly affect the structure of the accretion flow. This effect was first identified in Aql~X-1, where the intense radiation released during a burst was found to cool the hot inner flow and temporarily modify its geometry \citep{MaccaroneCoppi2003,Chen2012}, leading to a reduction of the Comptonized emission and a transient reconfiguration of the disc--corona system \citep[see also, e.g.][]{Fragile2020,Guver2022}. These observations demonstrate that thermonuclear bursts can modulate both the inflow and the outflow on short timescales, providing a direct probe of the interplay between accretion and energy release at the NS surface.

Type-I bursts also provide key constraints on the NS structure. During PRE bursts, the photospheric expansion phase is regulated by the Eddington luminosity, which depends on the NS mass and radius as well as on the atmospheric composition \citep{Guver2012}. By combining measurements of the touchdown flux and the apparent emitting area during the cooling tail, it is possible to place constraints on the NS mass–radius relation, and hence on the equation of state of ultra-dense matter \citep{Suleimanov2011,Suleimanov2017,Nattila2017}. Although systematic uncertainties remain, burst spectroscopy continues to represent one of the most powerful observational tools for probing the fundamental properties of NSs.
Finally, coherent oscillations observed at the NS rotation frequency \citep{Chakrabarty2003} during some X-ray bursts \citep{Strohmayer1996,Strohmayer2006} imply the presence of asymmetric temperature pattern at the NS surface \citep{Watts2012} and can also be used for constraining the NS equation of state \citep{Bhattacharyya2005,Kini2024}. 

\subsection{Boundary layer and spreading layer in NS LMXBs}\label{sec:bl}


In NS LMXBs, a substantial fraction of the accretion power is released in the boundary layer \citep{Sunyaev1986}, where the Keplerian accretion flow must decelerate and adjust to the NS surface. 
Unlike black holes, where matter can advect through the event horizon, accreting material in NSs is forced to radiate its kinetic energy, making the boundary layer a fundamental site of energy dissipation \cite{Popham2001}.

The accreted matter does not settle locally at the equator, but instead spreads over the NS surface, forming a belt (spreading layer) whose latitudinal extent increases with mass accretion rate \citep{Inogamov1999}. 
As the accretion rate rises, the spreading layer covers a larger fraction of the stellar surface and becomes increasingly dominated by radiation pressure. 
This geometry naturally leads to a situation in which the emitting area increases with accretion rate while the local flux remains close to the Eddington limit.

The presence of a boundary/spreading layer has important implications for the energetics of accretion onto NSs. 
In particular, up to $\sim 50\%$ of the total accretion luminosity can be released in this region \citep{Sunyaev1986,Sibgatullin2000}, making it a key component of the spectral energy distribution of NS XRBs. 
Spectral decomposition studies have shown that NS emission can often be described as a combination of a disc component and a boundary-layer component with distinct spectral properties. 
The latter is typically associated with a Comptonized or quasi-blackbody emission with a relatively stable temperature, consistent with radiation emerging near the local Eddington limit \citep[e.g.,][]{Gilfanov2003,Revnivtsev2006,Revnivtsev2013}. 
At high accretion rates, particularly in luminous Z-sources, this behaviour can be interpreted as evidence for an increase in the emitting area of the spreading layer \citep{Lin2009,Church2012}.

Thus, the accretion flow can be described as a combination of an optically thick accretion disc, an optically thin Comptonizing region (often referred to as a corona or hot inner flow), and a boundary layer connecting the inner disc to the NS surface. The observed X-ray spectrum reflects the contribution of these components: the disc produces a multicolour blackbody dominating the soft X-ray band, while Comptonization in the hot electron population generates a hard X-ray tail that can extend up to hundreds of keV (\citealt{Mitsuda1984}, but cf. \citealt{White1988}). The boundary/spreading layer contributes either as a quasi-blackbody component with characteristic temperatures of $\sim1$--$2$~keV or as the primary source of seed photons for Comptonization \citep{DoneGierlinski2003, Marino2022}.
The relative importance of these components evolves with accretion rate, giving rise to the canonical spectral states of NS systems: a hard state dominated by Comptonized emission from the hot inner flow, and a soft state in which the disc and the boundary/spreading layer dominate the radiative output \citep[e.g.,][]{Lin2007}.

The boundary/spreading layer is also a key contributor to the rapid variability observed in NS systems. Frequency-resolved spectroscopy has demonstrated that the fastest variability is associated with emission originating close to the NS surface, indicating that a large fraction of the accretion power is dissipated in the boundary layer \citep{Gilfanov2003,Revnivtsev2006}. These results support a picture in which the inner accretion flow and the NS surface are tightly coupled.

More recent spectral-timing studies have further highlighted that the inner accretion flow cannot be described as a set of independent, sharply separated regions. Analyses combining spectral and timing information \citep[e.g.,][]{Marino2023} indicate that the observed variability arises from a dynamically coupled system involving the accretion disc, a Comptonizing inner flow, and the boundary/spreading layer. In this framework, the transition between these components is gradual, and the dissipation of accretion power occurs over an extended region rather than at a well-defined interface. This reinforces the view that the boundary/spreading layer should be considered as part of a broader, continuously evolving inner accretion flow.

The presence of a spreading layer also has important implications for other accretion-related phenomena. In particular, the local accretion rate per unit area, which depends on how matter spreads over the NS surface, directly affects the ignition conditions for thermonuclear bursts \citep{Inogamov2010}. 
As a result, the geometry and evolution of the spreading layer play a crucial role not only in the release of accretion power, but also in the coupling between accretion and nuclear burning on the NS surface.

Overall, the boundary/spreading layer provides a key physical link between the accretion flow and the NS surface, and represents one of the fundamental differences between accretion onto NSs and black holes \citep{Cavecchi2026}.

\section{Accretion-ejection coupling in NS LMXBs}
\subsection{Jets in NS LMXBs}
\label{sec:jets}

Jets in NS low-mass X-ray binaries (NS LMXBs) have been extensively studied over the past two decades, with observations revealing a complex interplay between accretion processes, jet launching, and high-energy phenomena.  
Pioneering studies using Very Long Baseline Interferometry (VLBI) provided some of the earliest high-resolution images of jet structures in Sco~X-1, arguably the best studied NS LMXB \citep{Fomalont2001a,Fomalont2001b}.
These works studied the properties of the moving radio lobes with apparent relativistic speeds of 0.32--0.57$c$, corresponding to the two-sided radio jets launched from near the NS. 
This established a strong similarity between the jets from NSs and those from black hole systems. 
It was only 25 years later that \citet{FenderMotta2025}, using a sample of BH and NS binaries, found evidence that although jets from NSs are observationally similar to those from BHs, they are on average slower.

\cite{Fomalont2001a,Fomalont2001b} also discovered the presence of transient outflows that travel down the jet funnels producing consecutive brightenings in the lobes. 
Such outflows, later named Ultra Relativistic Flows (URFs) \citep{Motta2019}, are likely unique to NSs, as they have not been observed in any confirmed BH system as of yet, but have been identified in the NS LMXB Cir~X-1, and the enigmatic source SS433, which hosts a compact object the nature of which is still unclear. 
\cite{Fender2004} demonstrated that the NS Cir~X-1 produces URFs associated with transient radio flaring and extended nebulae powered by jet activity. 
These observations emphasized for the first time that jets from NSs can be powerful and morphologically complex, and capable of producing large-scale effects  \cite[see also][]{Cowie2025}. 
Based on high angular resolution X-ray observations with \textit{Chandra}, \cite{Migliari2005} showed that sequential brightenings of resolved knots along the jet axis in SS\,433 are consistent with the propagation of URFs along the jet, showing that such flows may affect the jet emission from the radio up to the X-rays. 

\cite{Migliari2006} presented the first comprehensive radio and X-ray monitoring of multiple NS LMXBs and revealed systematic differences in jet properties compared to BHs. 
Differently from BHs, NS systems are consistently radio-quiet relative to equivalent X-ray luminosities, with radio emission generally fainter by a factor of up to 30, and showing a substantially wider scatter in their radio/X-ray luminosity ratios than BHs \citep{Gallo2018,vandenEijnden2021}.
 The degree to which NS jets are quenched in the soft state is a more subtle issue than originally thought. Early claims that radio emission persists into the soft state in NS LMXBs relied heavily on Ser X-1 and 4U 1820--30 \citep{Migliari2004}, taken as evidence for weaker or absent soft-state quenching relative to BHs. However, the Ser X-1 detection was only marginal ($<4\sigma$) and lay along a sidelobe streak of a brighter nearby source; deeper, more recent observations \citep{Pattie2024} show that Ser X-1 radio emission is only rarely detectable in the soft state, and never as bright as originally reported. With this revision, 4U 1820--30 now appears to be an unusual outlier rather than representative of NS LMXBs as a class: with this possible exception, essentially all NS systems studied to date show a substantial downturn in radio emission in the soft state, even if it remains unclear whether this downturn is quantitatively as strong as in BHs \citep{Tudose2009, Miller-Jones2010}. Whether the residual, weak emission detected from some sources in very deep soft-state observations reflects incomplete quenching, a different jet-launching mechanism, or simply lower intrinsic jet power in NSs -- possibly related to the presence of a solid surface and different magnetic field configurations -- remains an open and important question.
\citet{Migliari2006} also presented the first evidence of coupling between accretion states and jet presence, with steady jets in low-luminosity hard states and discrete transient ejections in outbursts. 
Such a connection was later largely confirmed by other authors, including \cite{MunozDarias2014} who showed how the accretion-jet coupling is substantially the same in BH and NS systems (when low surface magnetic field NSs are considered). 

Thanks to new sensitive radio interferometers such as MeerKAT, one of the SKA precursors, it has become possible to detect faint radio jets in NS LMXBs in the active state \citep{Gasealahwe2023, Gasealahwe2024}, and to place deep upper limits on their radio emission in quiescence \citep{vandenEijnden2022}.
Such observations constrained jet radio luminosities at very low accretion rates.


Optical, near-infrared, and mm/submm observations provide a complementary window on jets in NS LMXBs, as they probe synchrotron emission originating much closer to the jet launching region than radio wavelengths. 
In the compact, self-absorbed jet scenario, the flat or slightly inverted synchrotron spectrum observed at radio frequencies is expected to break at higher frequencies, marking the transition from optically thick to optically thin emission. 
Above this jet spectral break, the emission steepens to a typical spectral index of $\alpha \sim -0.7$ ($F_\nu \propto \nu^{\alpha}$). 
This transition is thought to occur either close to the jet base, where the optical depth drops below unity (e.g. \citealt{Blandford1979}), or farther downstream at sites of particle acceleration associated with shocks along the jet (e.g. \citealt{Markoff2005,Marscher2008,Polko2014}). X-ray polarimetric measurements with \textit{IXPE} are now starting to probe this region directly, offering new constraints on the accretion flow and possibly the magnetic field geometry near the jet base (see Sect.~\ref{sec:ixpe}).
Using mid-infrared observations, \citet{Migliari2010} reported the first measurement of a jet spectral break in a NS system, 4U~0614+091, demonstrating that compact, steady jets in NS LMXBs produce broadband synchrotron spectra analogous to those observed in BH systems, albeit with systematically lower jet power. 
Subsequent optical and infrared studies revealed that the jet contribution to the spectral energy distribution is strongly state dependent. 
In particular, \citet{Baglio2016b} observed the disappearance of the infrared excess associated with optically thin synchrotron emission during soft accretion states for the NS LMXB 1RXS J180408.9$-$342058, providing direct evidence for jet suppression at high accretion rates in NSs. 
Owing to the limited availability of strictly simultaneous multiwavelength coverage (especially in the sub-mm/mm and far- to mid-infrared bands), the jet spectral break has so far been constrained in only a small number of NS XRBs, spanning a remarkably wide range of frequencies from $\sim10^{10}$ to $\sim10^{15}$~Hz \citep{DiazTrigo2018}.
More recently, \citet{Koljonen2025} reported the detection of a jet spectral break in the NS XRB 4FGL~J0427.8$-$6704, further expanding the still limited sample. 
It remains unclear whether jet spectral breaks in NS systems differ systematically from those observed in BH XRBs. In BH systems, jet breaks have been reported to cluster around $\nu_{\rm br}\sim10^{14}$ Hz during hard accretion states when compact jets are present (e.g., \citealt{Russell2013,Koljonen2015}); however, this claim should be treated with caution. Estimating $\nu_{\rm br}$ in BH systems typically requires decomposing the broadband SED into jet and thermal-disc components, assuming a negligible optical/IR synchrotron contribution from other components, and correcting for often uncertain interstellar reddening, all of which carry substantial systematic uncertainty; the inferred clustering is also in tension with the $L_{\rm radio}\propto L_{\rm X}^{0.7}$ jet-disc coupling relation, which would require considerable fine-tuning to accommodate a universal $\nu_{\rm br}$. By contrast, in 4U 0614+091 the accretion disc is sufficiently weak and the jet break frequency low enough that the break is directly and unambiguously visible in the SED, without requiring such a decomposition. Given these caveats, claims of a strong clustering of $\nu_{\rm br}$ around $10^{14}$ Hz in BH systems should be regarded as tentative rather than established.

\begin{figure}
    \centering
    \includegraphics[ width=\linewidth]{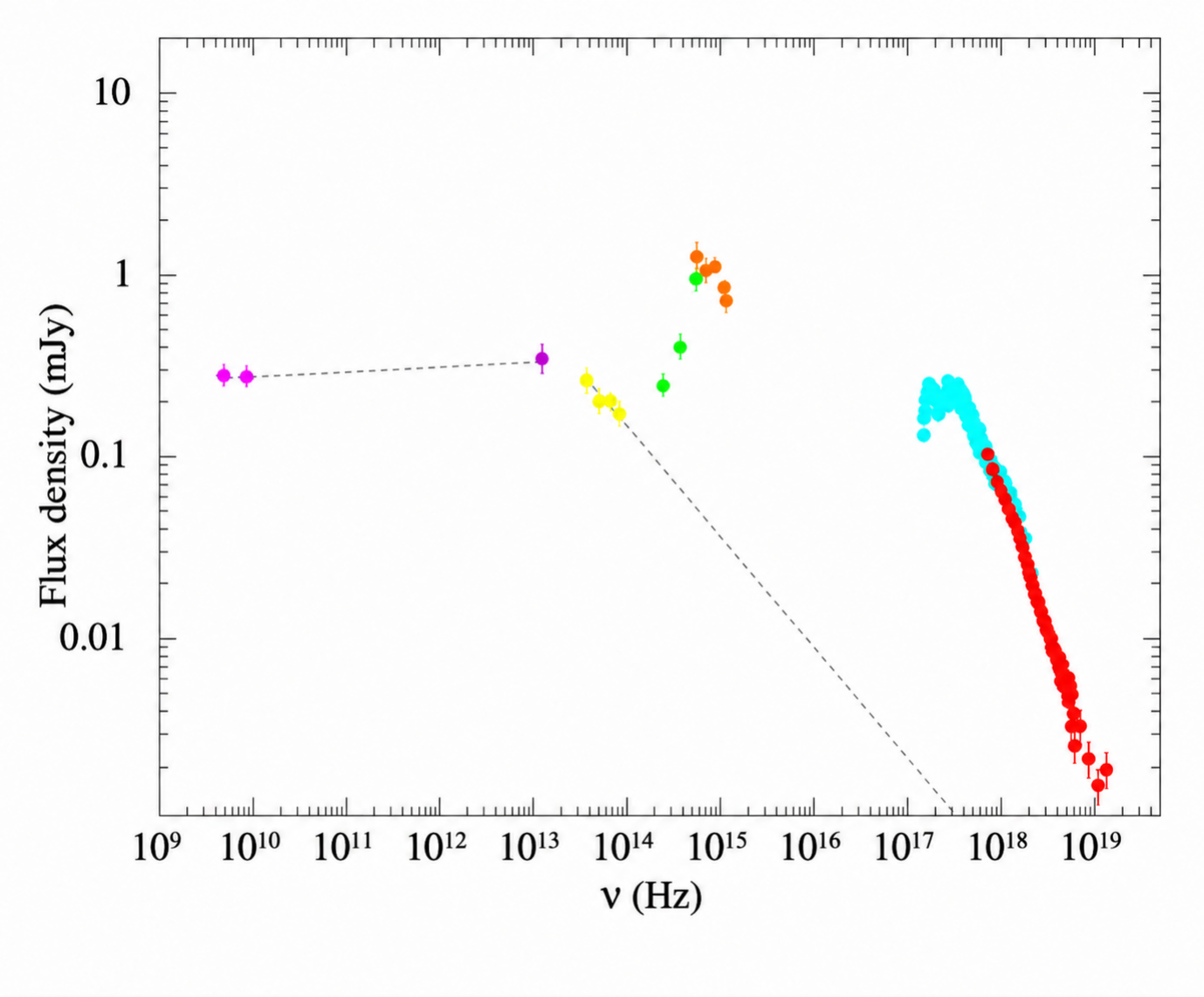}
    \caption{Multiwavelength observations of the NS XRB 4U 0614+091: VLA (magenta) Spitzer–MIPS (purple), \textit{Spitzer}–IRAC (yellow), SMARTS (green), \textit{Swift}–UVOT (orange), \textit{Swift}–XRT (light blue), and \textit{RXTE} (red), showing the flat optically thick and the break to the optically thin synchrotron emission from the compact jet (dashed lines). Fluxes are de-reddened. Adapted from \cite{Migliari2010}.}
    \label{fig:migliari}
\end{figure}

Beyond spectral diagnostics, fast optical and near-infrared timing offers a powerful probe of the physical coupling between accretion inflow and jet outflow. Such studies are observationally challenging for NS LMXBs because of the intrinsic faintness of their jets at opt/NIR wavelengths.
 
Despite this, \citet{Vincentelli2023} reported one of the most promising results to date, performing simultaneous near-infrared (HAWK-I@VLT) and X-ray (\textit{XMM–Newton}) observations of the NS LMXB 4U~1728$-$34. 
Achieving a near-infrared time resolution of $\sim120$~ms, they constructed a cross-correlation function between the two bands and detected a clear positive correlation, with the infrared leading the X-rays by $\approx30$--40~ms. 
Such a lead disfavours a pure jet origin of the fast infrared variability, which the authors instead attribute to synchrotron emission from an extended hot flow, while any jet-related lag is constrained to $\lesssim60$~ms. 
In a broader context, \citet{Vincentelli2023Nat} showed that the fast multiwavelength variability observed in both BH and NS binaries at high accretion rates can be explained by a shared accretion instability, supporting the idea that the physical processes governing the inner accretion flow and the associated ejections are largely independent of the nature of the compact object.

An additional key insight emerging from recent multiwavelength timing studies is that, in NS systems, the coupling between accretion inflow and jet outflow may be mediated by the boundary layer between the accretion disc and the NS surface (Sect.~\ref{sec:bl}). 
In particular, \citet{Marino2023} studied the NS LMXB 4U~1820$-$30 using coordinated X-ray and radio observations and showed that the variations of the jet radio emission are most naturally associated with changes in the boundary layer, rather than with hard-to-soft state transitions as in BHs. 
This interpretation is supported by the X-ray spectral evolution, in which the size of the Comptonisation seed-photon region, consistent with the boundary layer, changes between the high and low X-ray modes of the source. 
Similar conclusions were suggested by \citet{Russell2021}, who showed that the compact jet of 4U~1820$-$30 is quenched in its high X-ray modes, with the transition linked to the accretion rate rather than to the X-ray spectral hardness, suggesting that the boundary layer may play a central role in regulating the transfer of accretion power to the jet. 
These results provide compelling evidence that, while the overall accretion–jet coupling mechanism may be shared between BHs and NSs, the presence of a solid surface and boundary layer in NSs introduces an additional physical channel through which jet power can be modulated.

\subsection{Winds in NS LMXBs} 
\label{sec:winds} 

In addition to relativistic jets discussed in the previous section, accretion onto compact objects is also commonly associated with slower, disc-driven mass outflows, first proposed in the context of accretion disc corona sources by \citet{BegelmanMcKee1983}. In NS LMXBs, these outflows are observed as photoionized plasmas and are generally classified according to their ionization state, rather than temperature, into high-ionization and low-ionization winds \citep{DiazTrigo2006} (see also \citealt{MunozDarias2026} in this collection for a recent review).

High-ionization winds are primarily detected through X-ray spectroscopy as narrow absorption lines from highly ionized species (e.g. Fe\,\textsc{xxv} and Fe\,\textsc{xxvi}), often showing blueshifts that directly trace outflowing material. 
Based on current observations, $\sim13$ NS LMXBs show confident detections of such winds, while an additional nine sources have tentative detections \citep{MunozDarias2026}. 
These winds are typically associated with soft accretion states and are most frequently observed in high-inclination systems, suggesting an equatorial geometry \citep{Ponti2012,DiazTrigoBoirin2016}. 
This inclination dependence indicates that high-ionization winds may be a common, and possibly ubiquitous, feature of NS LMXBs, but are preferentially detected when the line of sight intercepts the outflow.

\begin{figure}
    \centering
    \includegraphics[width=1\linewidth]{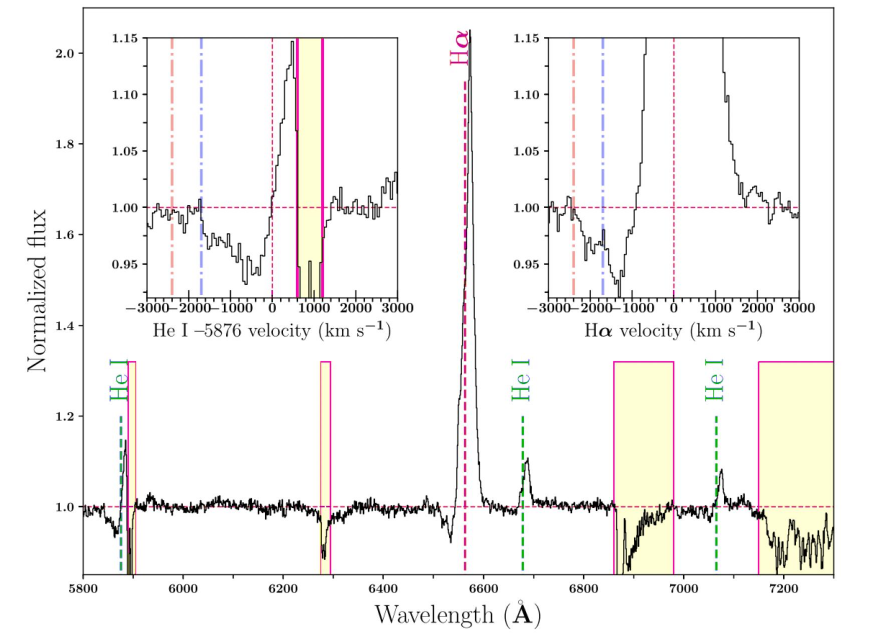}
    \caption{Optical spectrum of the NS LMXB Swift~J1858.6$-$0814 during a phase in which strong, high-velocity blueshifted absorption features are present, indicative of an outflowing accretion-disc wind. The inset panels show zoom-ins of the He\,\textsc{i}~$\lambda5876$ (left) and H$\alpha$ (right) lines in velocity space. The blue and red dashed-dotted lines mark the corresponding blue-edge velocities of the wind, reaching $\sim1700$ and $\sim2400$~km~s$^{-1}$, respectively. Yellow shaded regions indicate spectral intervals affected by interstellar or telluric contamination. Adapted from \cite{MunozDarias2020}.}
    \label{fig:winds}
\end{figure}

Low-ionization winds, on the other hand, are identified at optical and near-infrared wavelengths through P-Cygni profiles or emission-line asymmetries (see Fig.~\ref{fig:winds}), tracing cooler and less ionized material. 
These have been reported in a smaller number of systems, with two confident detections and three tentative cases, according to the most recent review by \citet{MunozDarias2026}. 
Unlike high-ionization winds, low-ionization components are often observed in harder accretion states, although their presence does not appear to be strictly limited to a single spectral regime.

An important aspect emerging from recent studies is that disc winds in NS LMXBs are likely multi-phase in nature. 
Observations of multiple absorption components with different ionization parameters and velocities \citep{Kallman2009,Xiang2009}, as well as the simultaneous detection of high- and low-ionization signatures in the same system \citep{MunozdariasPonti2022}, indicate that the outflow may consist of co-spatial or stratified plasma phases. 
This suggests that the wind structure is more complex than a single homogeneous component and evolves together with the accretion flow.

Furthermore, the phenomenology of disc winds is broader than that inferred from classical high-inclination dipping sources alone. Luminous NS systems such as GX 13+1 appear to host persistent, possibly multi-zone winds over a wide range of accretion states, consistent with thermal and radiative driving in large discs \citep{Allen2018}. Recent radio polarimetry adds further evidence that GX 13+1 is an unusually complex system \citep{Kashyap2025_GX13p1}, while an equivalent radio-polarimetric measurement of GX 17+2 provided a jet position angle -- inferred from the radio polarization angle rather than from a resolved jet image -- consistent with the X-ray polarization angle reported in Sect.~5.1 \citep{Kashyap2025_GX17p2}. In contrast, very high-inclination accretion disc corona (ADC) systems (e.g. 2S~0921$-$63 and 4U~1822$-$37) may be observed predominantly through scattered and reprocessed emission, rather than direct absorption, implying that wind signatures depend not only on their intrinsic properties but also on the viewing geometry \citep{Tomaru2020,Anitra2021}.

The presence of winds has important implications for the accretion process. 
Unlike disc atmospheres, true winds carry away mass and angular momentum, and in some systems the inferred mass outflow rate is comparable to or exceeds the mass accretion rate \citep{Neilsen2011,Ponti2012,Tetarenko2018}. 
This indicates that disc winds can significantly regulate the effective accretion rate onto the NS and influence the structure of the inner flow.

Several physical mechanisms have been proposed to drive these winds, including thermal pressure due to Compton heating, radiation pressure, and magnetohydrodynamic processes \citep{BegelmanMcKee1983,Eggum1988,Fukumura2010,Higginbottom2015,Ohsuga2011}. The relative importance of these mechanisms likely depends on luminosity, ionization state, and disc size, and may differ between high- and low-ionization components. Moreover, the presence of winds does not necessarily imply suppression of relativistic jets in NS systems: observational evidence suggests that winds and jets can coexist, particularly at high luminosities, pointing to a complex interplay between different outflow channels \citep{MunozDarias2020}.

Accretion disc winds are therefore a common feature of NS LMXBs and play a key role in shaping the accretion--ejection coupling, the observed spectral states, and the long-term evolution of these systems. The recent launch of \textit{XRISM}, with its high-resolution microcalorimeter spectrometer {Resolve}, has already begun to have a substantial impact on this field, providing direct measurements of wind velocity structure and kinematics that were previously inaccessible with CCD-resolution spectroscopy. Here we note that radiation scattered in an equatorial wind is polarized \citep{Nitindala2025} and can be used to determine orientation of the object on the sky (see Sect.~\ref{sec:polarisation}).

\section{Polarization of NS LMXBs} 
\label{sec:polarisation}

Polarimetric studies of NS LMXBs have historically been sparse, limited by the sensitivity of available instruments. 
Optical and near-infrared observations of systems such as Her~X-1 and Sco~X-1 revealed low levels of linear polarization, often interpreted as arising from Thomson scattering in asymmetric accretion geometries or from synchrotron emission in compact jets \citep[e.g.,][]{Egonsson1991,Schultz2004,Shahbaz2008,Russell2008}. 
These detections were typically at the sub-percent level and showed variability associated with orbital phase or source state transitions, offering indirect constraints on the structure and orientation of the accretion flow at large distances from a compact object.


The recent launch of \textit{IXPE} has opened a transformative window into the polarimetry studies in the X-ray range, where most of the released energy is radiated away in accreting BH and NS XRBs (see, e.g., \citealt{Dovciak2024,Ursini2024,Poutanen2024}).    
\textit{IXPE} allows to study the geometry of the accretion flows and magnetic fields in the vicinity of compact objects which is otherwise impossible to constrain.


\subsection{X-ray polarization and models}
\label{sec:ixpe}

Polarization in the X-rays was first studied with the OSO-8 satellite in the late 1970s. 
Marginally significant detections were obtained for NS-LMXBs Sco X-1 \citep{Long1979} and Cyg X-2 \citep{Weisskopf1976,Long1980} (see Table~\ref{tab:multiband_pol}).  
More recently, polarization at a $\sim$4\% level was detected from Sco X-1 with the PolarLight \citep{Long2022} with the polarization angle (PA) consistent with the direction of the relativistic jet. 

\begin{figure}
\centering
\includegraphics[width=0.5\linewidth]{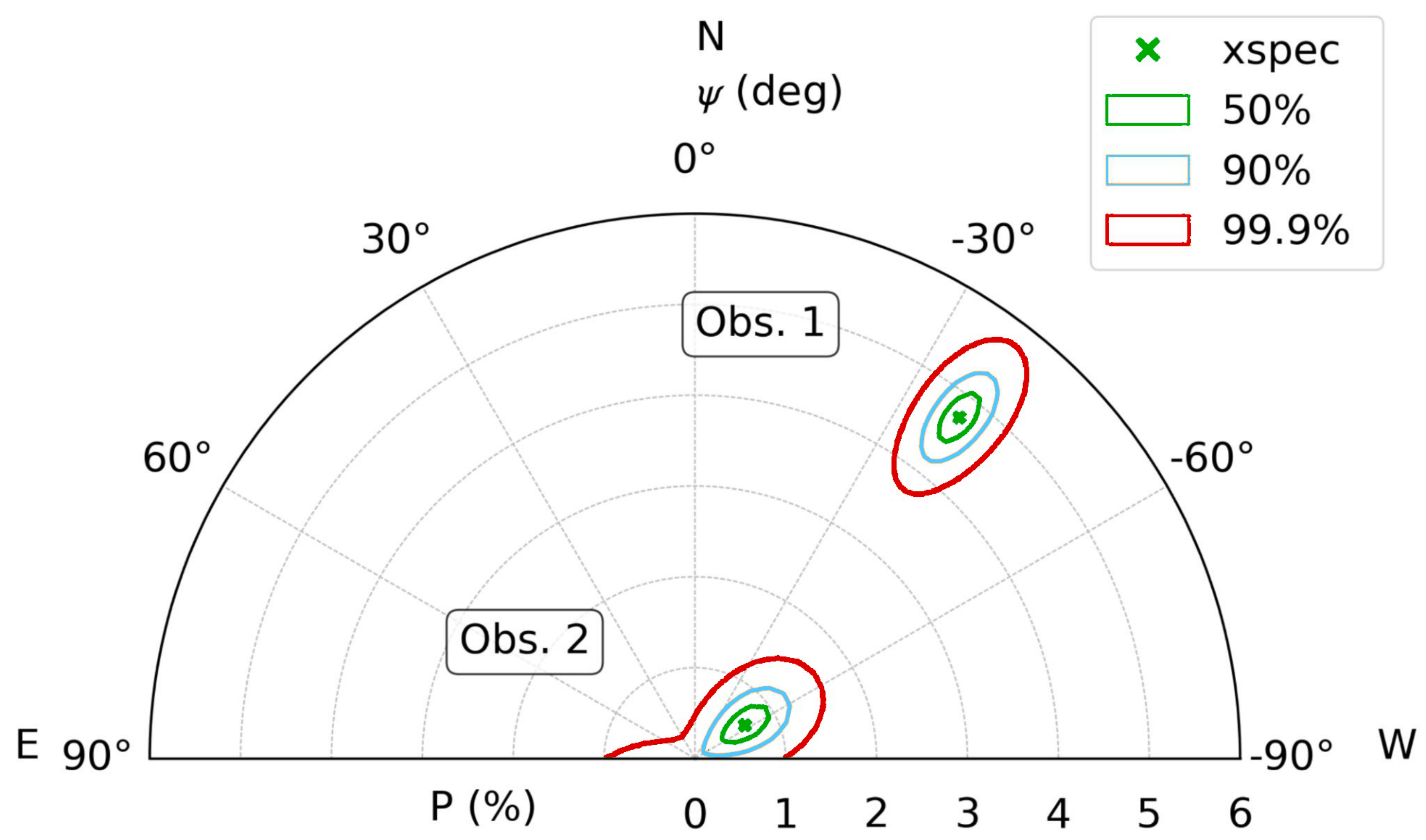}
\includegraphics[width=0.45\linewidth]{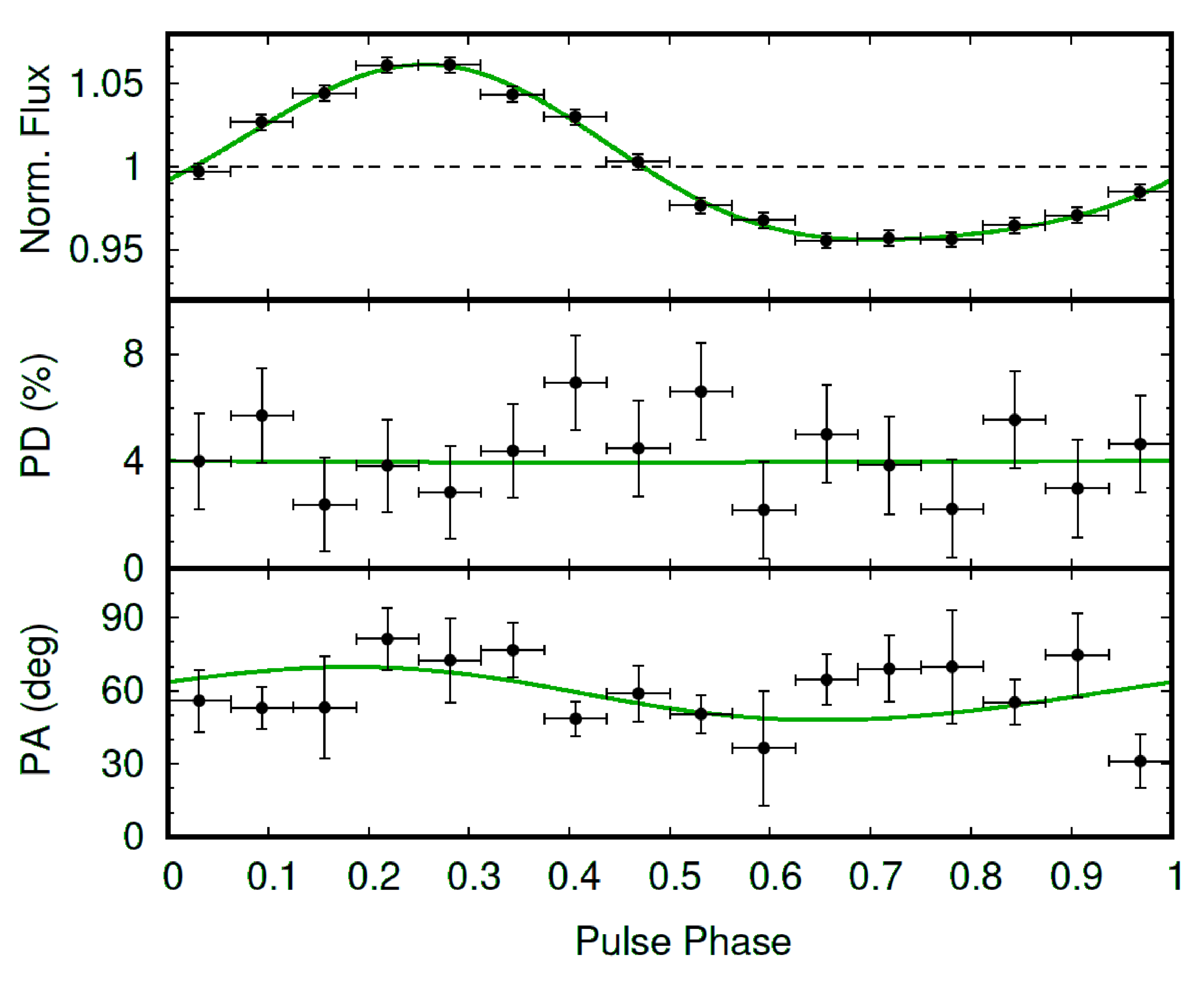}
\caption{Left: X-ray polarization properties of XTE~J1701$-$462 in the HB (Obs.~1) and NB  
\citep[Obs.~2; adapted from][]{Cocchi2023}. Right: Pulse phase-resolved X-ray polarization 
of AMXP SRGA~J144459.2$-$604207 (adapted from \citealt{Papitto2025}).}
\label{fig:xraypol}
\end{figure}

The first LMXB observed with \textit{IXPE} was an atoll source GS~1826$-$238 where no polarization was detected with a 3$\sigma$ upper limit of $<1.3\%$ \citep{Capitanio2023} implying an upper limit on the inclination of $i\lesssim50^\circ$ for most plausible geometries. In the Z-source Cyg~X-2, polarization was detected at a level of 1.8\% (see Table~\ref{tab:multiband_pol}, \citealt{Farinelli2023}) in the NB. The PA of $140^\circ\pm4^\circ$ was found to be consistent with the previous detections by OSO-8 and the position angle of the radio jet at $141^\circ$.
Such a coincidence immediately rejects simple models attributing polarization to an accretion disc or a boundary layer coplanar with the disc (both predicting PA perpendicular to the disc axis; e.g. \citealt{Cha60,Loktev2022}), but instead favours the vertically extended spreading layer \citep{Bobrikova2025}. This agreement between the polarization angle and the jet position angle is not, however, evidence that the jet itself is the source of the observed X-ray polarization: at these energies synchrotron emission from the compact jet contributes negligibly to the total X-ray flux (Sect.~\ref{sec:jets}), and so cannot plausibly drive a polarization signal of several percent. Instead, the coincidence is understood geometrically -- the jet and the disc/spreading-layer symmetry axis share a common orientation, so any structure aligned with that axis will naturally show the same PA, even though only the disc and spreading-layer material contribute the observed polarized flux. Two mechanisms remain viable for producing polarization along this shared axis: reflection of spreading-layer radiation from the disc \citep{Lapidus1985}, or scattering of that radiation in the wind \citep{Sunyaev1985,Tomaru2024,Nitindala2025}. The increase of the PD with energy, together with a high-energy (boundary/spreading-layer) PD of up to $\sim$4\% -- exceeding the $\sim$1.5\% expected from the spreading layer alone \citep{Bobrikova2025} -- favours one of these two additional-scattering scenarios over direct spreading-layer emission. Subsequent observations of Cyg~X-2 support this picture: in the HB, PD rose to 4.5\% with a PA of $128\degr\pm2\degr$ \citep{Gnarini2026}, consistent with the earlier OSO-8 measurement \citep{Long1980}. The same disc/jet-aligned geometry appears more broadly among Z-sources: GX 17+2 also showed a PA consistent with its jet position angle -- itself inferred from radio polarimetry rather than from a resolved jet image -- \citep{Kashyap2025_GX17p2}, and other Z-sources show a similar state-dependent PD, at 4--5\% level in the HB, 1--2\% in the NB and $\sim$2\% in the FB (\citealt{Cocchi2023,Fabiani2024,Gnarini2025}; see Fig.~\ref{fig:xraypol}, left).

Several atoll sources were observed by \textit{IXPE} in the soft state, resulting in a detection of polarization at a level of 1--3\%.
Some sources show a hint of increasing PD with energy \citep{Ursini2023,Saade2024} to a level of even 10\% and rotation of the PA by $\sim$90\degr\ \citep{DiMarco2023} which can be interpreted as evidence for perpendicular polarization of a softer accretion disc and a harder spreading layer. 
In other sources, only a low upper limit of $\sim$2\% could be established \citep{Capitanio2023,Gnarini2024b,Diaz2026}. 
And in GX~9+1 polarization was  detected only in the 2--3 keV band \citep{Tarana2025,ShyamPrakash2025}. 
The only hard-state atoll source was 4U\,1728$-$34 which showed PD of $\sim$2\% \citep{Kashyap2025_4U1728-34}. 

Some of the NS LMXBs showed unexpected variations not only in the PD but also in the PA. 
The PA observed from \mbox{Sco X-1} 2023 deviated from the PA observed by PolarLight and the jet direction by $\sim$46\degr\ \citep{LaMonaca2024} which implies significant variation in the orientation of the emission region and possibly jet direction. 
In \mbox{Cir X-1}, the PA showed variation by $\sim$50\degr\ along the orbit and by 67\degr\ when the data were divided by the hardness ratio \citep{Rankin2024}. 
This result was interpreted as evidence for a misalignment between the symmetry axes of the accretion disc and the Comptonizing region associated with the spreading layer caused by the misalignment of the NS's angular momentum with respect to the orbital one. 
Large variations of polarimetric properties were also observed in \mbox{GX~13+1}, a peculiar source with a unique position between atolls and Z-sources, with the PD varying from non-detectable to 5\% and the PA changing by 70\degr\ \citep{bobrikova2024a,Bobrikova2024b,DiMarco2025}. 
Observations covering its periodic dip revealed a PD of up to $\sim$9\% at the dip centre and a PA rotation of $\sim$60\degr\ between dip and off-dip states, constraining the geometry of an extended scattering medium (an oblate ADC or a disc wind) \citep{DiMarco2026}.
Strong variations in the PA could be attributed to the misalignment of the NS rotation axis, or to the presence of an asymmetric wind \citep{DiMarco2025,DiMarco2026} which scatters radiation from the central source \citep{Nitindala2025}.   
Further evidence for misalignment comes from a difference by 40\degr\ in the PA of the disc and hard (BL/SL) component in \mbox{GX 340+0} \citep{LaMonaca2024,LaMonaca2025}.

A distinct polarimetric regime is observed in accretion-disc corona (ADC) sources, i.e. high-inclination systems in which the direct view of the inner accretion flow is blocked by the outer disc, so that the observer receives only radiation scattered or reprocessed in an extended corona above the disc plane (see Sect.~\ref{Canonical_NS}). 
Because the observed flux is dominated by scattered light, ADC sources are expected to show substantially higher polarization degrees than typical NS LMXBs viewed at lower inclinations, with the polarization properties encoding the geometry of the scattering medium rather than of the emitting region itself.
Radiation polarized at 9\% level was observed from the ADC source 2S~0921$-$630 \citep{Tomaru2026}, implying a geometry similar to Seyfert 2 galaxies where polarization is produced by electron scattering. 
Similarly high polarization degrees have recently been reported for the eclipsing ADC source AX~J1745.6$-$2901, with PD reaching $\sim$15\% on average and $\sim$34\% during eclipse \citep{Mikusincova2025}, and for 4U~1822$-$37, with PD\,$\approx$\,8\% \citep{Anitra2026}, both consistent with scattering-dominated emission in a high-inclination geometry.

Among the AMXPs, only two sources were observed by \textit{IXPE}. 
In 2024, the transient SRGA~J144459.2$-$604207 was discovered by \textit{SRG}/ART-XC \citep{Molkov2024} with millisecond pulsations detected by NICER \citep{Ng2024}. 
Polarization with the PD of 2--6\% discovered by \citet{Papitto2025} (see Fig.~\ref{fig:xraypol}, right) could originate from electron scattering in an accretion shock close to the magnetic pole \citep{Viironen2004,Bobrikova2023}. 
A weak dependence of the PD and PA on pulse phase implies a hotspot geometry close to the rotational pole that is consistent with the small amplitude of pulsations \citep{Poutanen2006}. 
A similar picture emerged for SAX~J1808.4$-$3658 during its 2025 outburst: \citet{Papitto2026} detected a PD of $2.9\pm0.5\%$ (2--8 keV) with no significant energy or pulse-phase dependence, again pointing to hotspots closely aligned with the spin axis. The X-ray PA ($46\degr\pm5\degr$) is broadly consistent with the optical one \citep{Baglio2020}. They also reported a tentative ($2.5\sigma$) PD of $\sim16\%$ at the peak of a type-I X-ray burst, with a PA roughly orthogonal to that of the persistent emission.
 
\begin{figure}
\centering
\includegraphics[width=1\linewidth]{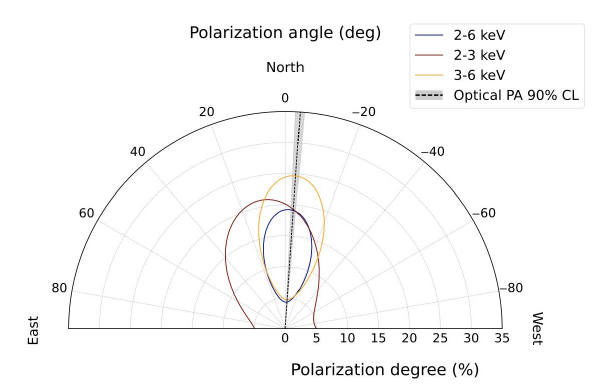}
\caption{Protractor plot of X-ray PD and PA for the high mode of the tMSP 
PSR\,J1023+0038 measured over the energy ranges 2--6, 2--3, and 3--6 keV, and displayed 
along the radial and azimuthal directions, respectively. Contours represent 90\% confidence 
regions. The black dashed line and gray shaded area indicate the optical R-band PA and its 
90\% c.l. uncertainty. Adapted from \citet{BaglioIXPE}.}
\label{fig:J1023xrays}
\end{figure}

X-ray polarimetric observations of the transitional millisecond pulsar (Sect.~\ref{tmsp_sec}) PSR\,J1023+0038 with \textit{IXPE}, reported in \citet{BaglioIXPE}, have provided the first constraints on the geometry of the high-energy emission in the sub-luminous disc state. 
A statistically significant linear polarization was detected during the high X-ray mode (see Sect.~\ref{tmsp_sec} for details), with a clear energy dependence: in the $2$--$6$ keV band the polarization degree reaches $12 \pm 3\%$ (see Fig.~\ref{fig:J1023xrays}), while it decreases when integrated over the broader $2$--$8$ keV range. 
The PA remains stable within uncertainties. 
In contrast, only upper limits were obtained during the low and flaring modes. 
The detection of a relatively high PD disfavours a purely thermal origin of the X-ray emission (e.g., disc blackbody or boundary-layer radiation), and instead supports a scenario in which a substantial fraction of the hard X-rays arise from synchrotron processes in a magnetically ordered region. 
The stability of the PA across observing intervals suggests that the large-scale emission geometry does not undergo major reconfigurations during mode switching.
Furthermore, the PA is not observed to vary with either orbital or pulse phase. 
In the framework of recent modelling by \cite{Veledina2025}, this behaviour is consistent with a predominantly poloidal magnetic-field configuration in the interaction region between the innermost accretion flow and the pulsar wind. 
Overall, these results indicate that, in the sub-luminous disc state, the X-ray emission is closely linked to the magnetically mediated interaction between the accretion flow and the pulsar wind, likely tracing synchrotron radiation produced near the disc--magnetosphere boundary.

\subsection{Optical and infrared polarization} 
\label{sec:optir_pol}

Multi-wavelength polarimetry is a powerful tool to probe the physical conditions and geometry of NS XRBs, providing unique insights into emission mechanisms, magnetic field structures, and scattering environments. In particular, comparing X-ray and lower-frequency (optical-infrared - OIR) polarization has proven especially effective, as shown by recent observational campaigns, primarily on BH systems, but also on at least one NS source \citep{Krawczynski2022,Mastroserio2025,BaglioIXPE, Kravtsov2025}.

\begin{sidewaystable}[]
\centering
    \resizebox{\textheight}{!}{%
    \begin{tabular}{l|l|c|c|l|l|>{\raggedright\arraybackslash}p{3.5cm}}    \hline   \hline
Source   & Obs. Band   & PD (\%) & PA (deg) & Possible source & State & Reference \\
      \hline
4U~0614+091 & Opt ($R$) & $3\pm1$ &  & Jet/scattering & NA &\citet{Baglio2014} \\
Sco X-1& \multirow{2}{*}{NIR}       &  $0.93\pm0.05$    & $147\pm3$  & \multirow{2}{*}{Jet}             &  \multirow{2}{*}{NA} &\citet{Russell2008} \\
&         &    $1.14\pm0.06$        &  $129\pm3$ &              & & \citet{Shahbaz2008}\\
& X-rays (5.6~keV)       &  $1.3\pm0.4$       & $57\pm6$ &             &   NA &\citet{Long1979}\\
& X-rays (2--8~keV)       &  $4.3\pm0.8$       & $52.6\pm5.4$ & SL           &  SA &\citet{Long2022}\\
& X-rays (2--8~keV)       &  $1.0\pm0.2$       & $8\pm6$ & SL/Wind            &   SA &\citet{LaMonaca2024}\\
Her X-1     & Opt       & 0.1--0.2        &  & Disc scattering & X-ray pulsars & \citet{Egonsson1991,Schultz2004}\\
& X-rays (2--8 keV)       & $8.6\pm0.5$     &$62\pm2$     & Hotspot & X-ray pulsar & \citet{Egonsson1991}\\
Cyg X-2     & X-rays (2.6 keV)    & $4.9 \pm 1.8$ &$138 \pm 10$ &   &  NA &\citet{Weisskopf1976,Long1980}\\
            & X-rays (2--8 keV)    & $1.8 \pm 0.3$ &$140 \pm 4$ & SL & SA &\citet{Farinelli2023}\\
           & X-rays (2--8 keV)    & $4.5 \pm 0.3$ &$128 \pm 2$ & Wind, reflection  & HB &\citet{Gnarini2026}\\
Cir X-1& X-ray (2--8 keV) & $1.6\pm0.3$/$1.4\pm0.3$ & $37\pm5$/$-12\pm7$ & Misaligned SL/BL & SS & \citet{Rankin2024}\\
GS~1826$-$238 & X-rays (2--8 keV)    &  $<1.3$  & --  &    & SS  & \citet{Capitanio2023} \\
XTE J1701$-$462     & X-rays (2--8 keV)    & $4.6 \pm 0.4$ & Variable & Reflection  & HB & \citet{Cocchi2023}\\
           &     & $\sim 0.6$ &  & Reflection   & NB & \citet{Cocchi2023}\\
GX 5$-$1& X-rays (2--8 keV)    &  $3.7\pm0.4$  & $-9.7\pm 2.0$  &    & HB  & \citet{Fabiani2024} \\
      &     & $1.8\pm0.4$ & $-9 \pm 4$ & Comptonization & NB & \citet{Fabiani2024} \\
GX 340+0 & X-rays (2--8 keV)    &  $4.3\pm0.4$  & $36\pm 3$  &  Misaligned SL/BL  & HB  & \citet{LaMonaca2024} \\
      &     & $1.4\pm0.3$ & $37 \pm 5$ & Wind & NB & \citet{LaMonaca2025,Bhargava2026} \\
GX 349+2 & X-rays (2--8 keV)    & $1.1 \pm 0.3$   & $29\pm 7$  &      & NB &\citet{LaMonaca2025b,Kashyap2025_GX349p2}\\
GX 17+2 & X-rays (2--8 keV)    & $1.9 \pm 0.3$   & $11\pm 4$  &  Reflection    & NB &\citet{Kashyap2025_GX17p2} \\
GX 9+1& X-rays (2--3 keV)    & $2.5 \pm 1.0$   & $2\pm 11$  &  Disc   & Atoll, SS &\citet{Tarana2025,ShyamPrakash2025}\\
GX 13+1& X-rays (2--8 keV)    & 2--9 & $\sim-5$ (variable) & Asymmetric wind/ADC  & Atoll, SS  & \citet{bobrikova2024a,Bobrikova2024b,Kashyap2025_GX13p1,DiMarco2025,DiMarco2026}\\
4U~1820$-$303 & X-rays (4--7 keV)    &  $2.0\pm0.5$  & $-55\pm7$ &    & Atoll, SS  & \citet{DiMarco2023} \\
GX 9+9& X-rays (2--8 keV)    & $1.4\pm 0.3$   & $68\pm 6$   &  SL + reflection  & Atoll, bright SS  & \citet{Ursini2023} \\
GX 3+1& X-rays (2--8 keV)    &   $<1.3$&   &    &   & \citet{Gnarini2024b} \\
4U 1624$-$49     & X-rays (2--8 keV) & $3.1\pm 0.7$ & $81\pm 6$ & Comptonization & Atoll, banana   & \citet{Saade2024,Gnarini2024}\\
Ser~X-1& X-rays (2--8 keV)    & $0.8\pm 0.4$   & $-60\pm 15$   & & Atoll, banana  & \citet{Ursini2024b} \\
4U~1728$-$34 & X-rays (2--8 keV) & $1.8\pm 0.8$ & $-35\pm 12$ &  & Atoll, HS  & \citet{Kashyap2025_4U1728-34} \\
4U~1735$-$44  & X-rays (2--8 keV) & $1.4\pm 0.7$ & $-29\pm 14$ &  & Atoll  & \citet{Diaz2026} \\
2S 0921$-$630 & X-rays (2--8 keV) & $8.5\pm 1.6$ & $-46\pm 6$ & ADC &   & \citet{Tomaru2026} \\
AX~J1745.6$-$2901 & X-rays (2--8 keV) & $14.7\pm 4.0$ & $122\pm 8$ & Disc wind scattering & ADC, eclipsing & \citet{Mikusincova2025} \\
4U~1822$-$37 & X-rays (2--8 keV) & $7.9\pm 0.6$ & $-24\pm 2$ & ADC/extended corona & ADC, eclipsing & \citet{Anitra2026} \\
SAX J1808.4$-$3658 & Opt ($BVRI$)& 0.2--2.2 & $30\pm20$ & Jet & AMXP, HS &\citet{Baglio2020}\\
 & X-rays (2--8 keV) & $2.9\pm0.5$ & $46\pm5$ & Hotspots (accretion shock) & AMXP, outburst decay &\citet{Papitto2026}\\
 & X-rays (2--6 keV) & $16\pm6$ & $-44\pm11$ & Scattering in extended structure? & Type-I burst peak &\citet{Papitto2026}\\
SRGA~J144459.2$-$604207   & X-rays (2--8 keV) & $2.3\pm0.4$ & $59\pm6$ & Accretion shock & AMXP, HS &\citet{Papitto2025}\\
PSR J1023+0038 & Opt ($R$)   & $1.41 \pm 0.03$ &   & Pulsar synchrotron  & tMSP &\citet{Baglio1023flaring,BaglioIXPE}\\
                     & X-rays (3--6 keV) & $12\pm 3$& $-2\pm 9$ & Wind--disc boundary & &\citet{BaglioIXPE}\\
MAXI J1807+132 & Opt ($BVRI$) & 0.3--0.6  & $82\pm13$ & Scattering in disc/jet & IS &\citet{Rout2025}\\
       &  & 0.6--1 & $140\pm16$ & Shocks in the jet & HS/trans.\ to SS  &\citet{Rout2025}\\
\hline
\end{tabular}}
\caption{Polarimetric properties of accreting NSs in low-mass X-ray binaries.
HB: Horizontal Branch in the Z-track; NB: Normal Branch; SA: Soft Apex;
HS: Hard State; IS: Intermediate State; SS: Soft state; NA: not available;
SL: Spreading Layer; ADC: accretion-disc corona.}
    \label{tab:multiband_pol}
\end{sidewaystable}

Optical and infrared (OIR) polarimetry can reveal contributions from electron scattering in the accretion disc and surrounding medium, or from synchrotron emission, often associated with relativistic jets or the hot flow — a geometrically thick, optically thin, radiatively inefficient inner accretion region that replaces the standard geometrically thin disc below a truncation radius \citep{Veledina2013}. In the X-ray band, polarimetric observations (now possible with missions such as \textit{IXPE}; Sect.~\ref{sec:ixpe}) offer a direct probe of the innermost accretion flow and magnetic field topology close to the NS. By combining polarimetric measurements across different wavelengths, it becomes possible to disentangle emission components and investigate the interplay between the compact object, its accretion flow, and possible outflows.

In practice, however, polarimetric studies of NSXRBs are particularly challenging. These systems are typically fainter than BHXRBs at most frequencies, and the intrinsic polarization levels are expected to be low. This is closely connected to differences in jet production efficiency between the two classes. BHXRBs exhibit a well-established radio/X-ray correlation in the hard state, with $L_{\rm radio} \propto L_{\rm X}^{0.6-0.7}$ \citep{Gallo2003}, where the radio emission is associated with compact jets (Sect.~\ref{sec:jets}). NSXRBs also follow a radio/X-ray correlation, but with a shallower slope ($L_{\rm radio} \propto L_{\rm X}^{0.3-0.4}$) and a significantly lower normalization \citep{Migliari2005,Migliari2006}.

The offset between NS and BH systems can be understood in terms of their different radiative efficiencies. In NSXRBs, the presence of a solid surface implies that a substantial fraction of the accretion power is radiated away, primarily in the boundary layer and from the stellar surface, making $L_{\rm X}$ a relatively direct tracer of the mass-accretion rate. In contrast, BHXRBs lack a surface, and part of the accretion energy can be advected across the event horizon or channelled into outflows without being efficiently radiated; however, since BH accretion discs can extend down to smaller inner radii than the NS surface or boundary layer, it is not obvious that BH systems are, in fact, less radiatively efficient overall. Consequently, at a given $L_{\rm X}$, BH systems may be accreting at a lower mass-accretion rate than NSs. Observationally, this produces a systematic offset in the radio/X-ray plane, with NSXRBs being significantly fainter in the radio band than BHXRBs at the same X-ray luminosity. This difference is commonly interpreted as evidence that jet production is less efficient in NSs, although differing radiative efficiencies likely also play a role. An alternative, not mutually exclusive explanation may lie in the location of the jet spectral break itself (Sect.~\ref{sec:jets}): if BH jet breaks typically fall in the near-infrared while NS jet breaks occur at higher, mid-infrared frequencies, intrinsically comparable jets could still produce systematically weaker polarization signatures for NSs at fixed observing wavelength, independent of any genuine difference in jet power.

As a consequence, synchrotron emission from jets is expected to contribute less to the total flux in NSXRBs at infrared frequencies, leading to intrinsically lower polarization fractions compared to BH systems. Indeed, no radio polarization attributable to optically thick synchrotron emission from compact jets has yet been detected in a NSXRB.

\begin{figure}
\centering
\includegraphics[width=1\linewidth]{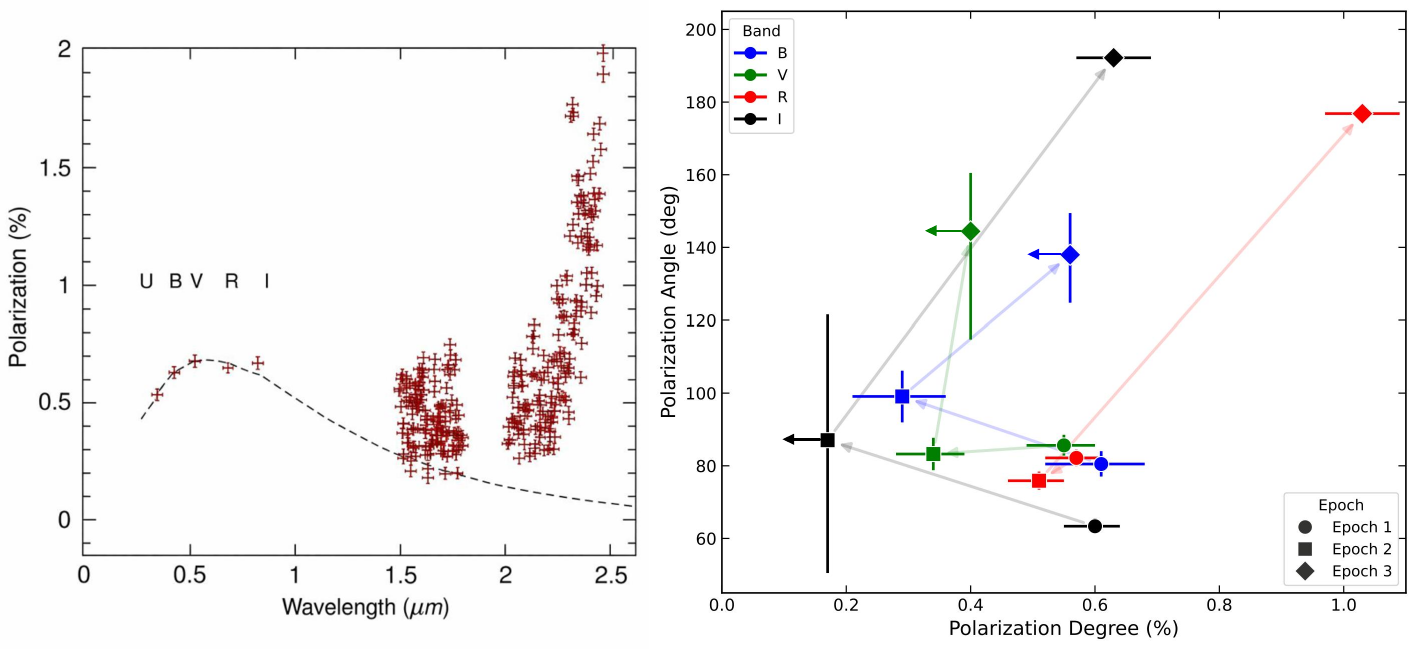}
\caption{\textit{Left}: optical and near-infrared polarization of Sco~X-1. The optical points (UBVRI bands) are taken from \cite{Schultz2004}. The dashed line is the interstellar polarization model fit to the optical; adapted from \cite{Shahbaz2008}. \textit{Right}: optical PA vs. PA of MAXI~J1807+132 for the three epochs. A clear rotation of the PA by approximately $\sim100\degr$ is observed in the R and I bands between the first two epochs and the third. Upper limits on the polarization level are reported at the $99.97\%$ credible interval, while all other uncertainties are quoted at the $\pm1\sigma$ credible level. Adapted from \cite{Rout2025}.}
\label{fig:scox1_maxi1807}
\end{figure}

Nevertheless, a few NS systems show hints of jet-related polarization at near-infrared and optical wavelengths. \cite{Russell2008} reported optical and infrared polarimetric signatures of relativistic jets in X-ray binaries at levels of $\sim$1--10\%, comparable to those observed in active galactic nuclei.   
A near-infrared $2.4\,\mu$m linear spectropolarimetry of persistent systems, including the NS sources Sco~X-1 and Cyg~X-2, was presented in \cite{Shahbaz2008} who detected polarization levels of $1.3\% \pm 0.10\%$ and $5.4\% \pm 0.7\%$, respectively. In the case of Sco~X-1, polarization was found to increase towards longer wavelengths (Fig.~\ref{fig:scox1_maxi1807}, \textit{left} panel), suggesting a contribution from synchrotron emission originating near the base of the jet.

Similarly, a $3\pm1\%$ linear polarization in the optical $r$-band for the persistent UCXB 4U~0614+091 was measured by \cite{Baglio2014}. Considering that the jet contributes only $\sim5\%$ of the total optical flux (from extrapolation of the optically thin jet spectrum; \citealt{Migliari2010,Baglio2014}), the inferred polarization implies a relatively high degree of magnetic-field ordering at the jet base. Using PD(\%)$=100\,f\,\frac{1-\alpha_{\rm thin}}{5/3-\alpha_{\rm thin}}$ \citep{Bjornsson1982,Russell2014}, the value of 15\% for the parameter $f$ describing the magnetic-field ordering was obtained by \citet{Baglio2014},  which is higher than typically inferred for both BH and NS systems.

It is important to note, however, that optical polarization may also arise from scattering within the accretion disc or its atmosphere. Therefore, while a jet origin remains plausible in several cases, a contribution from disc scattering cannot be excluded. Discriminating between these scenarios requires multi-band polarimetry to test the spectral dependence of the polarization and phase-resolved observations to search for orbital modulation, both of which provide critical constraints on the underlying emission mechanism.

\begin{figure}
\centering
\includegraphics[width=0.8\linewidth]{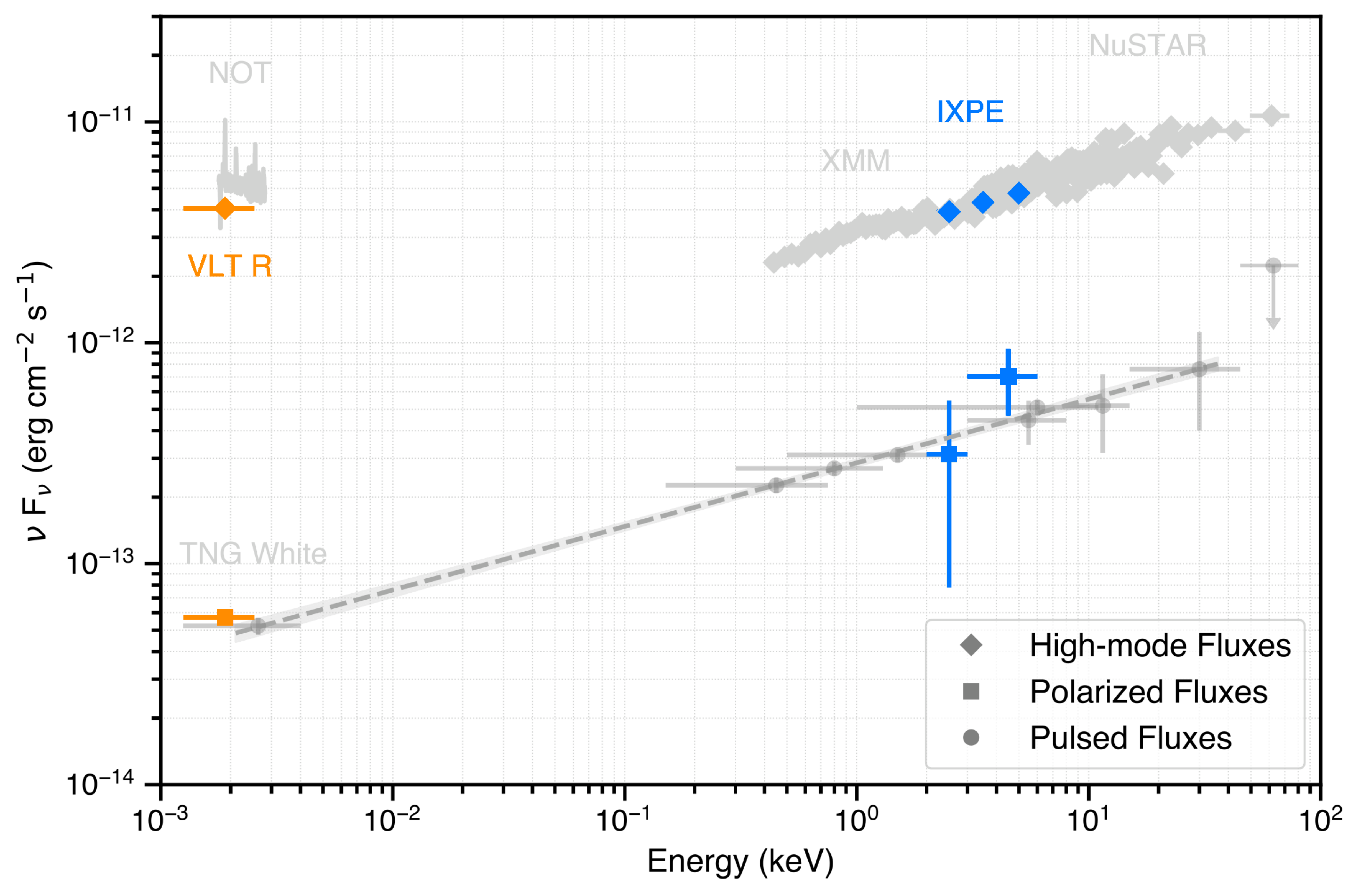}
    \caption{Spectral energy distribution of the optical/X-rays/ total (diamonds), pulsed (circles) and polarised (squares) emission of PSR J1023+0038 in the high mode of the sub-luminous disc state. Adapted from \cite{BaglioIXPE}.}
    \label{fig:SED_pol}
\end{figure}

More recently, \citet{Baglio2020} presented an optical polarimetric campaign of the AMXP SAX~J1808.4$–$3658 during its 2019 outburst. A low-frequency excess (red excess) in the spectral energy distribution, particularly in the z, i, and R bands, emerges as the outburst evolves, possibly indicating optically thin synchrotron emission from a jet, considering its simultaneity with radio detections of the rise of the compact jet. 
Polarimetric observations (BVRI bands; 440--768 nm) reveal variable LP ($0.2\%–2\%$), intrinsic to the source. The red LP spectrum supports a jet origin rather than scattering effects from the disc or propelled matter. The low-level, variable LP suggests the presence of strongly tangled magnetic fields near the base of the jet.

A recent study focused on the NS XRB MAXI~J1807+132 during an outburst \citep{Rout2025}, reporting optical polarization measurements obtained with the VLT over three epochs. The LP was low ($\lesssim 1\%$) during the first two epochs, which corresponds to the hard intermediate state, with its origin remaining unclear (possibly due to scattering in the disc, or synchrotron emission from the compact jet or hot flow). In the third epoch, taken just before an X-ray flare that marked the transition to the soft state, the LP increased to $\sim 1\%$ in the red bands, accompanied by a rotation of the polarization angle by $\sim 100\degr$ (Fig. \ref{fig:scox1_maxi1807}, \textit{right} panel). This behaviour may suggest that the LP in the third epoch is linked to shocks occurring in ballistic ejecta launched during the state transition.


Some NS XRBs, in contrast, show clear evidence of optical polarization due to scattering in the accretion disc. A good example is Her X-1 \citep{Egonsson1991,Schultz2004}, which exhibits an evolution of optical polarization over its orbital period, leading to the conclusion that it results from Thomson scattering by an optically thin material corotating with the system. 
\citet{Doroshenko2022} fitted the phase curves of the normalized Stokes parameters from \citet{Egonsson1991} using a Fourier series, obtaining the inclination angle of the orbital plane and the position angle of the orbital axis. Combined with new \textit{IXPE} X-ray polarimetry -- which independently measured a pulsar spin position angle of $57^\circ\pm2^\circ$ and a magnetic obliquity of $12^\circ\pm4^\circ$ -- this yielded a misalignment of at least $\sim20^\circ$ between the NS spin axis and the orbital angular momentum, providing direct polarimetric support for models in which the source's $\sim35$-day superorbital period is driven by free precession of the misaligned NS.
Further studies using spectro-polarimetry and time-resolved observations could provide additional constraints on the structure and evolution of the disc, as well as the mechanisms responsible for producing the observed polarization signatures.

Two transitional millisecond pulsars (Sect. \ref{tmsp_sec}) were observed with a polarimeter at optical frequencies ($BVRi$ bands, 440--793 nm).
XSS J12270$–$4859 was observed during its radio pulsar state \citep{Baglio2016}. No significant linear polarization was detected in any of the observed bands, with 3$\sigma$ upper limits as low as $1.4\% $ in the R-band, which placed stringent constraints on any polarized emission.
Polarimetric observation of PSR J1023+0038 were performed in the sub-luminous disc state (Sect.~\ref{tmsp_sec}) by several authors \citep{Baglio2016,Hakala2018,Baglio2023,Baglio1023flaring, BaglioIXPE}. Optical polarization at the $\sim1\%$ level was detected, with indications of significant variability that does not correlate with the orbital period. The most recent results suggest a possible increase in optical polarization during high modes \citep{Hakala2018,Baglio2023}. Furthermore, simultaneous optical and X-ray polarimetric observations have revealed that the polarized flux across both bands likely shares a common origin, attributed to synchrotron emission produced at the boundary between the pulsar wind and the inner accretion flow \citep{BaglioIXPE}. The same power-law possibly describes the pulsed and the polarized fluxes (Fig.~\ref{fig:SED_pol}), suggesting that the pulsations are highly polarized at all frequencies.

\section{Summary}

Fifty years after Nikolai Shakura and Rashid Sunyaev \citep{shakura1973} laid the theoretical foundations of disc accretion, neutron star low-mass X-ray binaries remain among the most productive laboratories in high-energy astrophysics. The framework they introduced -- angular-momentum transport through a viscous disc, gravitational energy release across the boundary between the Keplerian flow and the stellar surface -- has proven not merely durable but generative, providing the conceptual scaffolding on which five decades of observational discovery have been built.

This article has traced accretion from the outer disc to the stellar surface, following matter through the diverse phenomenology that makes NS LMXBs uniquely constraining. The zoo of systems reviewed in Section \ref{sec:flavours}  -- canonical atoll and Z-sources, ultracompact X-ray binaries, accreting and transitional millisecond pulsars -- reflects the wide range of accretion rates, magnetic field strengths, and evolutionary histories accessible within a single physical framework. The recycling of NSs into millisecond pulsars through sustained angular momentum transfer, confirmed observationally by the transitional pulsar systems that switch between accretion and rotation-powered states, stands as one of the clearest demonstrations that disc accretion shapes not only the immediate luminosity of a system but its long-term spin evolution.

The accretion disc itself encodes information about the inner-flow geometry in its spectral states and colour-colour diagrams, as established with EXOSAT and Ginga and refined with every subsequent generation of X-ray observatory. Timing diagnostics -- broadband noise, low- and high-frequency QPOs, and the kHz QPOs that probe orbital dynamics within a few gravitational radii of the NS -- established primarily with \textit{RXTE}, which also discovered coherent millisecond pulsations and burst oscillations, complement spectroscopy and together constrain the mass, radius, and spin of the underlying compact object. Relativistic reflection features and disc winds detected with XMM-Newton, Chandra, NuSTAR, and now \textit{XRISM} have revealed that accretion in NS LMXBs is as structurally rich as in their black-hole counterparts, while the presence of a solid surface and a magnetic field introduces physics that has no analogue in black-hole systems.

Thermonuclear Type-I X-ray bursts probe nuclear burning under extreme conditions, and the photospheric radius expansion events they produce remain among the most powerful tools for constraining the NS equation of state. The recent demonstration that bursts modulate the compact jet -- directly measuring, for the first time, the jet velocity in a NS system -- underlines the deep coupling between surface burning, inner accretion flow, and outflow, with future facilities such as \textit{NewAthena} \citep{Nandra2013,Cruise2025} expected to extend this diagnostic power considerably.

The boundary and spreading layers, where Keplerian motion decelerates onto the stellar surface, release up to half the total accretion luminosity and yet remain among the least directly constrained regions of the flow. Their spectral and timing signatures, and their role as a mediator between the disc and the jet, make them a central target for current and future observational efforts.

The jets and winds reviewed in Sections \ref{sec:jets} and \ref{sec:winds}  demonstrate that NS LMXBs participate fully in the disc–jet coupling seen across the accreting compact-object population, while differing from black-hole systems in jet power, quenching behaviour, and the additional physical channel provided by the boundary layer. Disc winds, detected across a wide range of ionisation states and accretion regimes, carry away mass and angular momentum at rates that can rival the accretion rate itself, shaping both the spectral evolution and the long-term history of these systems.

A transformative recent development has been the opening of the X-ray polarimetric window by \textit{IXPE}. For the first time, new, model-dependent constraints on the geometry of the emission regions (the disc, the corona, the boundary and spreading layers, the jets) are accessible, complementing the degeneracies that spectral modelling alone cannot resolve. Early \textit{IXPE} results have already confirmed theoretical predictions for the spreading layer geometry, revealed the alignment between polarisation angle and jet axis in several systems, and uncovered unexpected polarisation signatures in accreting millisecond pulsars that challenge simple geometric models. Optical and near-infrared polarimetry has provided complementary constraints on Thomson scattering geometries and jet contributions at large scales.

Over fifty years, the diverse phenomenology of NS LMXBs has converged into a coherent physical picture. The Shakura–Sunyaev accretion disc, the NS surface and its boundary layer, recycled millisecond pulsars, thermonuclear bursts, relativistic jets, the ionised wind, and the polarised X-ray emission are all manifestations of a single physical process: the infall of matter under gravity onto a rotating, magnetised, solid body, observed from different angles, at different epochs, and with ever more powerful instruments. The next fifty years, with facilities such as the next generation of X-ray timing and polarimetry missions, square-kilometre radio arrays, and gravitational wave detectors probing the NS interior directly, promise to be no less revelatory.





\bmhead{Acknowledgements}
The authors acknowledge the support of the International
Space Science Institute (ISSI) in Bern through ISSI Workshop “Accretion Disks: The First 50 Years”. M.C.B. acknowledges support from the INAF Fundamental Research Grant XBOOM.
A.P.N. and J.P. were supported by the Research Council of Finland Centre of Excellence in Neutron-Star Physics (grant 374064) and the Magnus Ehrnrooth Foundation (A.P.N.).
We thank Giulia Illiano for comments and fruitful discussion about accreting pulsars.
This work is dedicated to the memory of Maurizio Falanga, who was to have been a co-author of this review.

 \section*{Declarations}
\subsection*{Conflict of interest}
The authors declare no competing interests.

\bibliography{refs_arxiv}

\end{document}